\documentclass[notitlepage,aps,11pt,amsmath,amsfonts,amssymb,nofootinbib]{revtex4-1}
\usepackage{enumerate}
\usepackage{graphicx}

\begin{document}

\newcommand{\new}[1]{\ensuremath{\blacktriangleright}#1\ensuremath{\blacktriangleleft}}

\title{Finsler geometric extension of Einstein gravity}

\author{Christian Pfeifer}
\email{christian.pfeifer@desy.de}
\affiliation{II. Institut f\"ur Theoretische Physik und Zentrum f\"ur Mathematische Physik, Universit\"at Hamburg, Luruper Chaussee 149, 22761 Hamburg, Germany}
\author{Mattias N.\,R. Wohlfarth}
\email{mattias.wohlfarth@desy.de}
\affiliation{II. Institut f\"ur Theoretische Physik und Zentrum f\"ur Mathematische Physik, Universit\"at Hamburg, Luruper Chaussee 149, 22761 Hamburg, Germany}

\begin{abstract}
We construct gravitational dynamics for Finsler spacetimes in terms of an action integral on the unit tangent bundle. These spacetimes are generalizations of Lorentzian metric manifolds which satisfy necessary causality properties.  A coupling procedure for matter fields to Finsler gravity completes our new theory that consistently becomes equivalent to Einstein gravity in the limit of metric geometry. We provide a precise geometric definition of observers and their measurements, and show that the transformations by means of which different observers communicate form a groupoid that generalizes the usual Lorentz group. Moreover, we discuss the implementation of Finsler spacetime symmetries. We use our results to analyze a particular spacetime model that leads to Finsler geometric refinements of the linearized Schwarzschild solution. 
\end{abstract}

\maketitle

\section{Introduction}
The weak equivalence principle states that the trajectories of small test bodies, neither affected by gravitational tidal forces nor by forces other than gravity, are independent of their internal structure and composition~\cite{Will:2005va}. Experimentally, this principle is confirmed with extremely high precision~\cite{Bae[]ler:1999zz}; in gravity theory it has been implemented  already by Newton who postulated that gravitational mass should equal inertial mass, and then by Einstein who formulated the motion of test bodies in terms of geodesics on Lorentzian spacetime. These trajectories extremize the Lorentzian length integral which is interpreted physically as proper time. The fundamental geometric object entering this construction is the Lorentzian spacetime metric. This observation led Einstein to the development of a gravity theory that determines the metric and so provides a dynamical background geometry for point particles, observers and physical fields.

The essential point in Einstein's implementation of the weak equivalence principle by a clock postulate is the use of geometric concepts, not the particular choice of metric geometry. Indeed, more general geometries can be used which automatically realize this principle. Here we consider Finsler geometry~\cite{Chern,Bucataru} which generalizes metric geometry by providing a very general length functional for curves $\tau \mapsto \gamma(\tau)$ on a manifold $M$,
\begin{equation}\label{eq:Fact}
S[\gamma]=\int d\tau\, F(\gamma(\tau), \dot \gamma(\tau))\,.
\end{equation}
The Finsler function $F$ maps the points of the curve and the attached four-velocities into real numbers and is homogeneous of degree one in its second argument to ensure the reparametrization invariance of $S[\gamma]$. The usual length measure associated to a Lorentzian metric~$g$ is obtained for the special case $F(\gamma,\dot \gamma)=\sqrt{|g_{ab}(\gamma)\dot \gamma^a\dot \gamma^b|}$. Physically, we interpret the integral above as a generalized clock postulate and point particle action. 

On this basis we will develop a consistent gravity theory which determines the Finsler function dynamically. Our construction builds on the precise definition of physical Finsler spacetimes in~\cite{Pfeifer:2011tk}, where we established a minimal set of requirements on the function $F$ so that it can describe a geometric spacetime background suitable for physics. In particular, Finsler spacetimes provide a well-defined notion of causality and the possibility to formulate field theory actions. A generalization of gravity based on Finsler geometry has the potential to explain various issues that are not naturally explained by Einstein gravity. Indeed, it has been argued that Finsler geometry in principle can address the rotational curves of galaxies and the acceleration of the universe without introducing dark matter~\cite{Chang:2008yv} or dark energy~\cite{Chang:2009pa}, and that it admits sufficiently complex causal structures that allow a consistent geometric explanation of superluminal neutrino propagation~\cite{Pfeifer:2011ve}. Here we will find that Finsler gravity may also explain the fly-by anomaly~\cite{Lammerzahl:2006ex} in the solar system.

Our presentation is structured as follows. We will begin in section~\ref{sec:fstreview} with a brief review of our definition of Finsler spacetimes and of the mathematical tools and geometric objects needed. Moreover, we will define field theory action integrals on Finsler spacetimes. Equipped with these concepts we will show in section~\ref{sec:obs} how to model observers on Finsler spacetimes and discuss how they perform measurements. We will prove that two different observers are related by a transformation composed out of a certain parallel transport and a Lorentz transformation. The set of these transformations has the algebraic structure of a groupoid that reduces to the usual Lorentz group in the metric geometry limit. In section~\ref{sec:gravdyn} we will present our new theory of Finsler gravity, including a matter coupling principle, which geometrically extends Einstein gravity without introducing new fundamental scales. We will derive the Finsler gravity field equation by variation and prove that it reduces consistently to the Einstein field equations  in the metric limit. Symmetries of Finsler spacetimes will be introduced in section~\ref{sec:fsym}. The maximally symmetric solution of vacuum Finsler gravity turns out to be standard Minkowski spacetime. Section~\ref{sec:Corr} considers a spherically symmetric perturbation around this vacuum which is found to be a refinement of the Schwarzschild solution that shows how Finsler gravity could resolve the fly-by anomaly. We conclude in section~\ref{sec:disc}. Appendix~\ref{app:details} presents technical details of our derivations.

\section{Finsler spacetime geometry}\label{sec:fstreview}
The central idea behind the generalization of Einstein gravity presented in this article is the description of spacetime and its dynamics by Finsler geometry instead of metric geometry. In this section we will review the basic geometric concepts available on Finsler spacetimes~\cite{Pfeifer:2011tk}. These were introduced as a generalization of Lorentzian metric spacetimes; they allow full control of the null geometry and a clean definition of causality, which is essential for the description of light and observers. In particular, we will describe non-linear connections and curvature; moreover, we will explain how to obtain well-defined field theory actions.

\subsection{Basic concepts}\label{sec:basic}
The definition of Finsler spacetimes involves the eight-dimensional tangent bundle $TM$ which is the union of all tangent spaces to the underlying four-dimensional event manifold $M$. Thus, any point $P\in TM$ is a tangent vector to $M$ at some point $p\in M$; there is a natural projection ${\pi:TM\rightarrow M, P\mapsto p}$. It is convenient to use the so-called induced coordinates on the tangent bundle which are constructed as follows. Let $(x)$ be coordinates on some open neighbourhood $U\subset M$ of $p=\pi(P)$. With respect to these we can express $P = y^a \frac{\partial}{\partial x^a}_{|x(p)}$; the induced coordinates of $P$ are then $(x(p),y)$. The corresponding induced coordinate basis of $TTM$ will be denoted by ${\big\{\partial_a=\frac{\partial}{\partial x^a}, \bar\partial_a=\frac{\partial}{\partial y^a}\big\}}$ and that of its dual $T^*TM$ by $\{dx^a, dy^a\}$.  

\vspace{6pt}\noindent\textbf{Definition 1.}
\textit{A Finsler spacetime $(M,L,F)$ is a four-dimensional, connected, Hausdorff, paracompact, smooth manifold~$M$ equipped with a continuous function $L:TM\rightarrow\mathbb{R}$ on the tangent bundle which has the following properties:
\begin{enumerate}[(i)]
\item $L$ is smooth on the tangent bundle without the zero section $TM\setminus\{0\}$;\vspace{-6pt}
\item $L$ is positively homogeneous of real degree $n\ge 2$ with respect to the fibre coordinates of $TM$,
\begin{equation}\label{eqn:hom}
L(x,\lambda y)  = \lambda^n L(x,y) \quad \forall \lambda>0\,;
\end{equation}
\item \vspace{-6pt}$L$ is reversible in the sense 
\begin{equation}\label{eqn:rev} 
|L(x,-y)|=|L(x,y)|\,;
\end{equation}
\item \vspace{-6pt}the Hessian $g^L_{ab}$ of $L$ with respect to the fibre coordinates is non-degenerate on $TM\setminus A$ where~$A$ has measure zero and does not contain the null set $\{(x,y)\in TM\,|\,L(x,y)=0\}$,
\begin{equation}
g^L_{ab}(x,y) = \frac{1}{2}\bar\partial_a\bar\partial_b L\,;
\end{equation}
\item \vspace{-6pt}the unit timelike condition holds, i.e., for all $x\in M$ the set 
\begin{equation}
\Omega_x=\Big\{y\in T_xM\,\Big|\, |L(x,y)|=1\,,\;g^L_{ab}(x,y)\textrm{ has signature }(\epsilon,-\epsilon,-\epsilon,-\epsilon)\,,\, \epsilon=\frac{|L(x,y)|}{L(x,y)}\Big\}
\end{equation}
contains a non-empty closed connected component $S_x\subset \Omega_x\subset T_xM$.
\end{enumerate}
The Finsler function associated to $L$ is $F(x,y) = |L(x,y)|^{1/n}$ and the Finsler metric $g^F_{ab}=\frac{1}{2}\bar\partial_a \bar\partial_b F^2$.}

\begin{figure}[h]
\vspace*{12pt}
\includegraphics[width=0.3\textwidth]{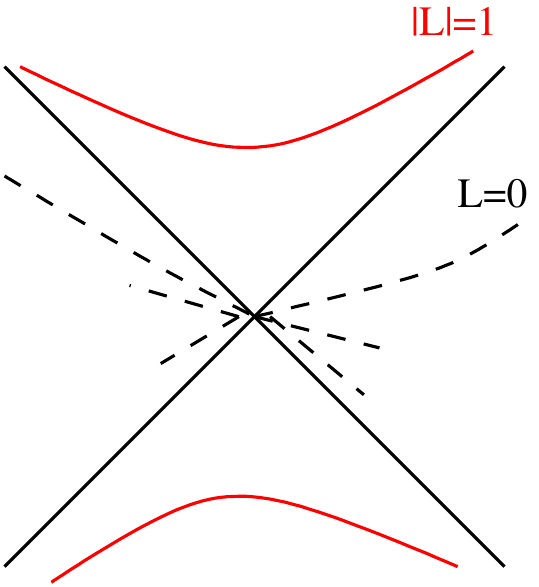}
\caption{\label{fig:nullstructure}\textit{Geometric structures implemented in every tangent space $T_xM$ by Definition~1 of Finsler spacetimes. The solid lines show the guaranteed cone of timelike vectors with the shells of unit timelike vectors and null boundary. The dotted lines indicate a potentially more complex null structure.}}
\end{figure}

Lorentzian metric spacetimes $(M,g)$ arise from Definition~1 in case $L(x,y)=g_{ab}(x)y^ay^b$; then $g^L = \epsilon\, g^F= g$ and $S_x$ is the set of unit $g$-timelike vectors. For general function $L$, the relation between 
the inverse Finsler metric $g^F$ and the inverse Lagrange metric $g^L$ is given by
\begin{equation}\label{eq:gftogl}
g^{F\,ab}=\frac{nL}{2 |L|^{2/n}}\Big(g^{L\,ab}+\frac{2(n-2)}{n(n-1)L}y^ay^b\Big).
\end{equation}
The definition of Finsler spacetimes guarantees a well-defined causal structure by the existence of timelike vectors that form an open convex cone with null boundary in every local tangent space, see figure~\ref{fig:nullstructure}. It provides full control of the geometry along the null directions where ${L(x,y)=0}$: we could show that the geometric concepts of connections and curvature that will be discussed below can be extended to this set, which is not possible in standard textbook formulations of Finsler spaces. Furthermore, by constructing a theory of electrodynamics on Finsler spacetimes we proved that light propagates along null directions. For further details see~\cite{Pfeifer:2011tk}. 

\begin{figure}[h]
\vspace{12pt}
\includegraphics[width=0.6\textwidth]{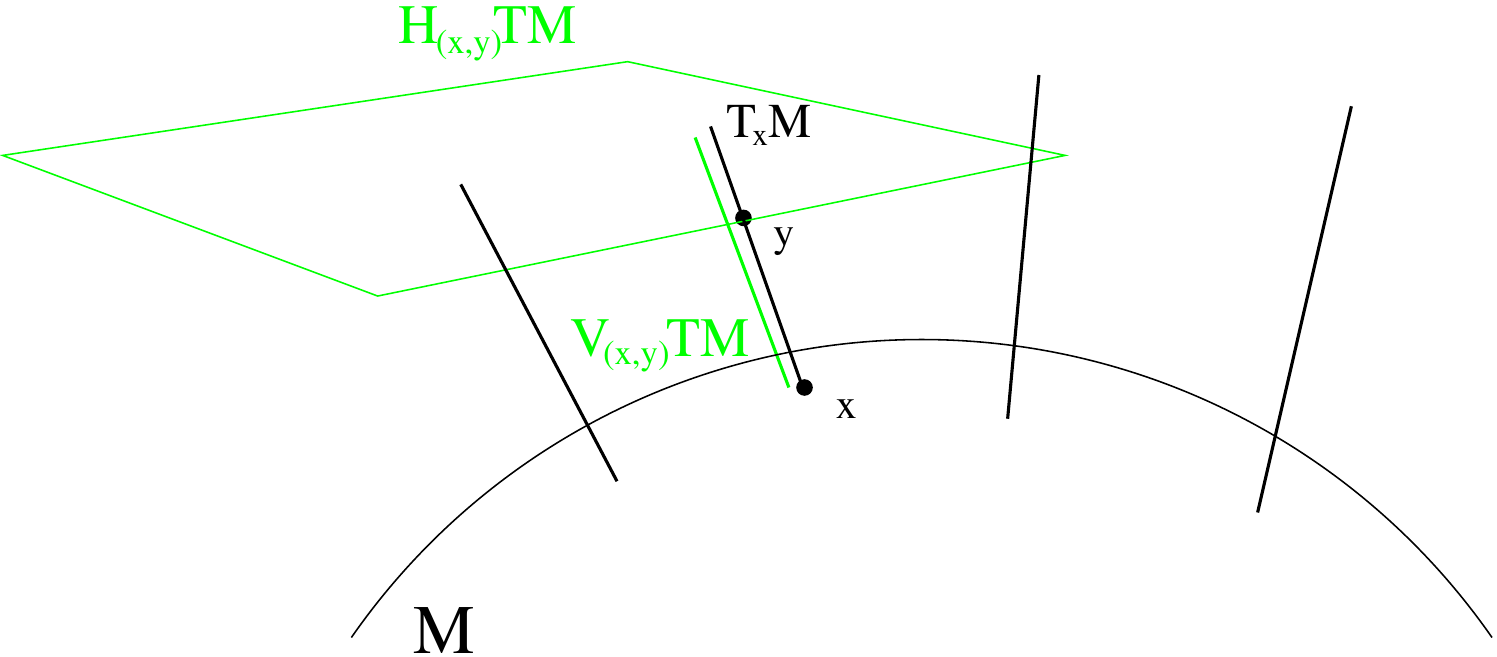}
\caption{\label{fig:verthori}\textit{Tangent bundle geometry: decomposition of $T_{(x,y)}TM$ into horizontal and vertical parts.}}
\end{figure}

The geometry of Finsler spacetimes is formulated by extending the standard language of Finsler geometry, see e.g.~\cite{Chern, Bucataru}; this is a special geometry on the tangent bundle  based on the Finsler function $F$ that appears in the clock postulate. The basic geometric object deduced from the Finsler function is the Cartan non-linear connection. Any non-linear connection on $TM$ is equivalent to a unique decomposition of all tangent spaces to $TM$ as $T_PTM=H_PTM\oplus V_PTM$, see figure~\ref{fig:verthori}. While the vertical bundle $VTM$ is canonically spanned by $\{\bar\partial_a\}$, the horizontal bundle $HTM$ is spanned by $\{\delta_a=\partial_a-N^b{}_a\bar\partial_b\}$ where the $N^a{}_b(x,y)$ are the coefficients of the non-linear connection. Then the dual bundle $(VTM)^*$ is spanned by $\{\delta y^a=dy^a+N^a{}_bdx^b\}$, and $(HTM)^*$ by $\{dx^a\}$. For the Cartan non-linear connection, the connection coefficients are determined by the fundamental functions $F$ or $L$ as 
\begin{equation}\label{eq:nonlin}
N^a{}_{b}=\frac{1}{4}\bar\partial_b\Big[g^{Faq}\big(y^p\partial_p\bar\partial_qF^2-\partial_qF^2\big)\Big]=\frac{1}{4}\bar\partial_b\Big[g^{Laq}\big(y^p\partial_p\bar\partial_qL-\partial_qL\big)\Big] .
\end{equation}
The equality of the two expressions for $N^a{}_b$ follows from the proof of Theorem~2 in~\cite{Pfeifer:2011tk} which makes particular use of the Euler theorem $y^a\bar\partial_a f(x,y) = m f(x,y)$ for $m$-homogeneous functions. Note that the right hand side is also valid on the null structure where $F$ is not even differentiable. 

Under tangent bundle changes of coordinates induced from coordinate transformations $(x)\rightarrow(\tilde x(x))$ of the manifold, the non-linear connection coefficients $N^a{}_b$ transform in such a way that the horizontal/vertical basis $\{\delta_a, \bar\partial_a\}$ of $TTM$ and the dual basis $\{dx^a, \delta y^a\}$ transform as vectors or one-forms over the manifold,
\begin{equation}
(\delta_a, \bar\partial_a)\rightarrow\Big(\frac{\partial \tilde x^b}{\partial x^a}\tilde\delta_b,\frac{\partial \tilde x^b}{\partial x^a}\tilde{\bar\partial}_b\Big),\quad 
(dx^a, \delta y^a)\rightarrow\Big(\frac{\partial x^a}{\partial \tilde x^b} d\tilde x^b,\frac{\partial x^a}{\partial \tilde x^b} \delta\tilde y^b\Big).
\end{equation}
Tensors on $TM$ that obey the standard tensor transformation law under manifold induced coordinate changes are called distinguished, or, in short, d-tensors. An example for a d-vector is the horizontal lift $X^H$ of a vector $X$ in $T_xM$ to $H_{(x,y)}TM$,
\begin{equation}\label{eq:horlift}
X=X^a\partial_a{}_{|x} \mapsto X^H=X^a\delta_a{}_{|(x,y)}\,.
\end{equation}
The inverse map is given by the pushforward along the projection, $\pi_*X^H = X$. The horizontal tangent spaces to $TM$ can thus be identified with the tangent spaces to the manifold $M$; this will be important for the identification of field components of tensors over the tangent bundle with field components measured over the spacetime manifold.

A d-tensor of particular importance for our construction of gravitational dynamics for Finsler spacetimes is the curvature of the Cartan non-linear connection which measures the integrability of the horizontal bundle $HTM$. It is defined as
\begin{equation}\label{eq:fcurv}
R^a{}_{bc}=[\delta_b,\delta_c]^a=-y^q\big(\delta_b\Gamma^{\delta a}{}_{qc}-\delta_c\Gamma^{\delta a}{}_{qb}+\Gamma^{\delta a}{}_{pb}\Gamma^{\delta p}{}_{qc}-\Gamma^{\delta a}{}_{pc}\Gamma^{\delta p}{}_{qb}\big),
\end{equation}
using as a short-hand notation the generalized Christoffel symbols 
\begin{equation}
\Gamma^{\delta\, a}{}_{bc}=\frac{1}{2}g^{F\,aq}\big(\delta_bg^F_{qc}+\delta_cg^F_{qb}-\delta_qg^F_{bc}\big)=\frac{1}{2}g^{L\,aq}\big(\delta_bg^L_{qc}+\delta_cg^L_{qb}-\delta_qg^L_{bc}\big)
\end{equation}
which are related to the non-linear connection coefficients via $N^a{}_b=\Gamma^{\delta a}{}_{bc}y^c$. The proof of the above equality is given in Theorem~2 of~\cite{Pfeifer:2011tk}. 

In order to formulate covariant differential equations in the language of Finsler geometry, a linear covariant derivative $\nabla$ that acts on tensor fields over $TM$ is needed. This must be compatible with the horizontal/vertical structure so that the identification of horizontal tangent spaces with tangent spaces to the manifold stays intact under transports. In the literature exist four such covariant derivatives with slightly different properties. We will employ the Cartan linear covariant derivative defined by
\begin{equation}\label{eq:cartanlin}
\nabla_{\delta_a}\delta_b=\Gamma^{\delta\, q}{}_{ab}\,\delta_q,\quad 
\nabla_{\delta_a}\bar\partial_b=\Gamma^{\delta\, q}{}_{ab}\,\bar\partial_q,\quad 
\nabla_{\bar\partial_a}\delta_b=\frac{1}{2}g^{F pq}\bar\partial_pg^F_{ab}\,\delta_q,\quad 
\nabla_{\bar\partial_a}\bar\partial_b=\frac{1}{2}g^{F pq}\bar\partial_pg^F_{ab}\,\bar\partial_q\,.
\end{equation}
In later sections we will often need the Cartan linear covariant derivative with respect to horizontal directions; we will then use the abbreviation $\nabla_a\equiv \nabla_{\delta_a}$. The action of $\nabla$ on the following Sasaki type metric on $TM$,
\begin{equation}\label{eq:sasaki}
G=-g^F_{ab}\,dx^a\otimes dx^b - \frac{1}{F^2}g^F_{ab}\,\delta y^a\otimes\delta y^b\,,
\end{equation}
gives zero. Hence this metric is covariantly constant with respect to the Cartan linear connection; it will play a crucial role as an integration measure on $TM$.

This completes our quick review of geometric objects needed in this article. For later use observe that all the objects definable from $F$, i.e., the Finsler metric $g^F$,the Cartan non-linear connection $N^a{}_b$, its curvature, and the Cartan linear connection, are invariant under the transformation $L\rightarrow L^k$. In the metric limit $L(x,y)=g_{ab}(x)y^ay^b$ they reduce to known constructions from metric geometry: the curvature essentially becomes the Riemann curvature tensor, $R^a{}_{bc}(x,y)=-R^a{}_{dbc}(x)y^d$; the generalized Christoffel symbols become the usual Christoffel symbols, $\Gamma^{\delta\, a}{}_{bc}(x,y)=\Gamma^{a}{}_{bc}(x)$; the non-linear connection now is a linear connection, $N^a{}_b(x,y) = \Gamma^a{}_{bc}(x)y^c$; the Cartan linear covariant derivative in horizontal directions becomes the Levi-Civita covariant derivative while it becomes trivial in vertical directions.

\subsection{Action integrals}\label{sec:int}
We saw that all geometric objects on Finsler spacetimes are homogeneous tensor fields on the tangent bundle; the same will be true for physical fields in our construction. The formulation of field theory actions needs the well-defined integration procedure developed in~\cite{Pfeifer:2011tk}. Since integrals over the tangent bundle over homogeneous functions diverge, we must consider integrals over the unit tangent bundle
\begin{equation}
\Sigma=\{(x,y)\in TM| F(x,y)=1\}\,.
\end{equation}
On this domain we have a natural volume element constructed from the pull back of the Sasaki-type metric~(\ref{eq:sasaki}). Thus, actions arise as integrals of scalar functions $f$ on $TM$ restricted to $\Sigma$ as~$f_{|\Sigma}$.

To perform such integrals explicitly, we change coordinates on $TM$ from the induced $Z^A=(x^a,y^a)$ to the more convenient coordinates $\hat Z^A= (\hat x^a(x), u^\alpha(x,y), R(x,y))$, where $\alpha=1\dots 3$, the $u^\alpha$ are zero-homogeneous, and $R=F(x,y)$. Now $\Sigma$ is defined by $R=1$ and described by coordinates $(\hat x,u)$.
The coordinate transformation matrices on $TM$ are
\begin{equation}
\frac{\partial \hat Z^A}{\partial Z^B} = \left[\begin{array}{c|c} \delta^a_b&0\\ \hline \begin{array}{c} \partial_b u^\alpha \\ \partial_b |L|^{1/r} \end{array}& \begin{array}{c} \bar\partial_b u^\alpha \\ \bar\partial_b |L|^{1/r} \end{array}\end{array}\right],\qquad
\frac{\partial Z^A}{\partial \hat Z^B} = \left[\begin{array}{c|c} \delta^a_b&\begin{array}{cc}0\quad& \;0 \end{array}\\ \hline \hat\partial_by^a& \begin{array}{cc} \partial_{u^\beta} y^a & \frac{y^a}{R} \end{array} \end{array}\right].
\end{equation}
and satisfy the invertibility properties
\begin{subequations}\label{eq:invvv}
\begin{equation}\label{eq:inv1}
\frac{\partial \hat Z^A}{\partial Z^C}\frac{\partial Z^C}{\partial\hat Z^B} = 
\left[\begin{array}{c|c}\delta^a_b & 0 \\ \hline \begin{array}{c} \partial_b u^\alpha+\bar\partial_c u^\alpha \hat\partial_b y^c \\ \partial_b |L|^{1/r} + \bar\partial_c|L|^{1/r} \hat\partial_b y^c \end{array}& \begin{array}{cc} \bar\partial_c u^\alpha \partial_{u^\beta} y^c & \bar\partial_c  u^\alpha \frac{y^c}{R} \\ \bar\partial_c|L|^{1/r} \partial_{u^\beta} y^c & 1 \end{array}\end{array}\right] = 
\left[\begin{array}{c|c}\delta^a_b & 0 \\ \hline \begin{array}{c}0 \\ 0 \end{array}& \begin{array}{cc} \delta^\alpha_\beta & 0 \\ 0 & 1 \end{array}\end{array}\right],
\end{equation}
\begin{equation}\label{eq:inv2}
\frac{\partial Z^A}{\partial \hat Z^C}\frac{\partial \hat Z^C}{\partial Z^B} = 
\left[\begin{array}{c|c}\delta^a_b & 0 \\ \hline  \hat \partial_b y^a + \partial_b u^\gamma\partial_{u^\gamma} y^a + \frac{y^a}{R} \partial_b |L|^{1/r} & \partial_{u^\gamma} y^a \bar\partial_b u^\gamma + \frac{y^a}{R} \bar\partial_b|L|^{1/r} \end{array}\right] = 
\left[\begin{array}{c|c}\delta^a_b & 0 \\ \hline 0& \delta^a_b \end{array}\right].
\end{equation}
\end{subequations}

We now calculate the pullback of the Sasaki type metric to $\Sigma$ in order to determine the relevant volume form. First we transform~(\ref{eq:sasaki}) to the new coordinates $(\hat x,u,R)$, which yields
\begin{equation}\label{eq:sasaki2}
G=-g^F_{ab}\,d\hat x^a\otimes d\hat x^b-\frac{1}{R^2}h^F_{\alpha\beta}\,\delta u^\alpha\otimes\delta u^\beta-\frac{1}{R^2}\,dR\otimes dR
\end{equation}
in terms of $h^F_{\alpha\beta}=g^F_{ab}\partial_\alpha y^a\partial_\beta y^b$ and $\delta u^\alpha= du^\alpha + (\bar\partial_b u^\alpha N^b{}_a - \partial_a u^\alpha) d\hat x^a$. Then the pull-back to $\Sigma$ with $R=1$ becomes
\begin{equation}
G^*=-g^F_{ab}{}_{|\Sigma}\,d\hat x^a\otimes d\hat x^b- h^F_{\alpha\beta}{}_{|\Sigma}\,\delta u^\alpha\otimes\delta u^\beta.
\end{equation}

Using the shorthand notation $g^F = |\textrm{det }g^F_{ab}|$ and $h^F = |\textrm{det }h^F_{\alpha\beta}|$, a well-defined integral over $\Sigma$ of a homogeneous tangent bundle function $f$ now reads
\begin{equation}
\int_\Sigma d^4\hat x d^3u\ \sqrt{g^Fh^F}{}_{|\Sigma}\, f(x,y)_{|\Sigma}\,.
\end{equation}
For tangent bundle functions $A^a(x,y)$ that are homogeneous of degree $m$ the following formulae for integration by parts hold
\begin{subequations}\label{eq:intbp}
\begin{eqnarray}
\int_\Sigma d^4\hat x d^3u\ \sqrt{g^Fh^F}{}_{|\Sigma}\big(\delta_a A^a\big)_{|\Sigma} &=&- \int_\Sigma d^4\hat x d^3u\ \sqrt{g^Fh^F}{}_{|\Sigma}\Big[ \big(\Gamma^{\delta\,p}{}_{pa}+S_a\big)A^a\Big]_{|\Sigma}\,,\\
\int_\Sigma d^4\hat x d^3u\ \sqrt{g^Fh^F}{}_{|\Sigma}\big(\bar\partial_a A^a\big)_{|\Sigma} &=&- \int_\Sigma d^4\hat x d^3u\ \sqrt{g^Fh^F}{}_{|\Sigma}\Big[\big(g^{F\,pq}\bar\partial_ag^F_{pq} - (m+3) y^p g^F_{pa}\big)A^a\Big]_{|\Sigma}\,,
\end{eqnarray}
\end{subequations}
where $S^a{}_{bc} = \Gamma^{\delta\,a}{}_{bc}-\bar\partial_b N^a{}_b$ and $S_a = S^p{}_{pa}$.
These formulae can be proven with the help of the coordinate transformation relations~(\ref{eq:invvv}).

The definitions and mathematical techniques presented in this section are the foundation of the following new developments. In the next section we present how observers are modelled and perform measurements before we turn to the construction of a gravity theory for Finsler spacetimes.

\section{Observers and measurements}\label{sec:obs}

In order to study physics on general Finsler spacetimes it is necessary to define a mathematical model of physical observers and to determine how they measure time, spatial distances, and physical fields. Guided by general relativity, freely falling observers move on trajectories that extremize the proper time integral; variation of~(\ref{eq:Fact}) here leads to Finsler geodesics. Moreover a model of observers requires four tangent vectors that build an orthonormal frame; then measurable quantities are the components of physical fields with respect to this frame, evaluated at the observers position. We explicitly calculate the illustrative example of an observer's measurement of the speed of light on which the results of~\cite{Pfeifer:2011ve} are based. To compare measurements of different observers it is necessary to communicate the results obtained by one observer to another. This communication is realized by a certain class of transformations between different observers; we will show that these transformations have the algebraic structure of a groupoid that generalizes the usual Lorentz group in metric geometry. 

\subsection{Orthonormal observer frames}
An observer moves along a spacetime curve $\tau\mapsto \gamma(\tau)$ in $M$ with timelike tangents. The parametrization can be chosen so that $\dot\gamma \in S_\gamma$ is unit timelike; according to Definition~1 we now have $|L(\gamma,\dot\gamma)|=1$ and the signature of $g^L_{ab}(\gamma,\dot\gamma)$ is Lorentzian.\footnote{In the following we often use the very intuitive notation $(\gamma,\dot\gamma)$ for points of the tangent bundle $\dot\gamma\in T_\gamma M\subset TM$, which is analogous to the coordinate representation $(x(\gamma),y(\dot\gamma))$.} Then the clock postulate~(\ref{eq:Fact}) tells us that $\dot \gamma$ must be interpreted as the local unit time direction of the observer. We may write the normalization condition in the form $g^F_{(\gamma,\dot\gamma)}(e_0,e_0)=1$ using the horizontal lift $e_0=\dot\gamma^H$ of $\dot\gamma$, see~(\ref{eq:horlift}). 

To identify the three-space seen by an observer, we will complete $e_0$ to a four-dimensional basis $e_\mu$ of $H_{(\gamma,\dot\gamma)}TM$; as explained before the projections of the $e_\alpha$ for $\alpha=1... 3$ by $\pi_*$ into $T_\gamma M$ then are identified as the spatial tangent directions to the manifold. We determine the three horizontal vectors $e_\alpha$ by the condition $g^F_{(\gamma,\dot\gamma)}(e_0,e_\alpha)=0$. This construction is justified by the observation that a horizontal three-space is defined by a conormal horizontal one-form. The only linearly independent one-form available in terms of geometric data is the vertical form $dL = \bar\partial_a L\, \delta y^a$. This  can be mapped globally to the horizontal one-form $\widetilde{dL}=\bar\partial_a L\,dx^a$, which is proportional to the Cartan one-form known from Finsler geometry and is a Lagrangian analogue of the Poincar\'e one-form in Hamiltonian mechanics. The condition $\widetilde{dL}_{(\gamma,\dot\gamma)}(e_\alpha)=0$ on $e_\alpha$ is equivalent to that stated above in terms of the Finsler metric. We remark that the $e_\alpha$ may depend less trivially on $\dot\gamma=\pi_*e_0$ than in Lorentzian geometry because of their defining equation $g^F_{(\gamma, \dot\gamma)}( e_0, e_\alpha)=0$. 

The definition of unique unit directions in the three-dimensional span $\langle e_\alpha\rangle$ requires orthonormalization. For this purpose we use the Finsler metric to set $g^F_{(\gamma,\dot\gamma)}(e_\alpha,e_\beta)\sim \delta_{\alpha\beta}$, assuming definite signature. The choice of the metric $g^F$ for orthonormalization is preferred over that of $g^L$, since only $g^F$ is invariant under $L\rightarrow L^k$ as are the geometrical objects on Finsler spacetimes, see section~\ref{sec:basic}. We will now prove a useful theorem on the relation between the signatures of the metrics $g^L$ and $g^F$ where both are defined; a corollary will then confirm our assumption of a definite signature of the Finsler metric on $\langle e_\alpha\rangle$. 

\vspace{6pt}\noindent\textbf{Theorem 1.}
\textit{On the set $TM\setminus (A\cup \{L=0\})$ the metric $g^L$ is nondegenerate of signature $(-1_m,1_p)$ for natural numbers $m,p$ with $m+p=4$. Then the Finsler metric has the same signature where $L(x,y)>0$, and reversed signature $(-1_p,1_m)$ where $L(x,y)<0$.}

\vspace{6pt}\noindent The observer's time direction $e_0$ is in $S_\gamma$ which is contained in $TM\setminus (A\cup \{L=0\})$. This tells us that the metric $g^L_{(\gamma,\dot\gamma)}$ has signature $(-1_3,1_1)$ for $L(\gamma,\dot\gamma)>0$ and $(-1_1,1_3)$ for $L(\gamma,\dot\gamma)<0$. We also know that $g^F_{(\gamma,\dot\gamma)}(e_0,e_0)=1$. Hence we conclude from Theorem 1:

\vspace{6pt}\noindent\textbf{Corollary.}
\textit{The Finsler metric $g^F_{(\gamma,\dot\gamma)}$ evaluated at the tangent bundle position of an observer has Lorentzian signature $(-1_3,1_1)$, and the unit spatial directions satisfy $g^F_{(\gamma,\dot\gamma)}(e_\alpha,e_\beta)=-\delta_{\alpha\beta}$.}

\vspace{6pt}\noindent\textit{Proof of Theorem 1.} 
By the definition of Finsler spacetimes the metric $g^L$ is  non-degenerate on $TM\setminus A$, hence also on the smaller set excluding the null structure on which $g^F$ is defined. Now observe that if an inner product is given by a matrix $C_{ab} = A_{ab}+ B_aB_b$ and $A_{ab}$ has indefinite signature $(-1_m,1_p)$, then the signature of $C_{ab}$ is found to be $(-1_{m+1},1_{p-1})$ for $A^{-1\,ab}B_aB_b<-1$ and $(-1_m,1_p)$ for $A^{-1\,ab}B_aB_b>-1$; for $A^{-1\,ab}B_aB_b=-1$ the result is the once degenerate signature $(-1_{m-1},1_p,0_1)$. This can be seen in a Sylvester normal form basis for $A$ by using the remaining $SO(m,p)$ freedom. We can apply this result to our situation by identifying
\begin{equation}
A = \frac{nL}{2|L|^{2/n}}\,(g^L)^{-1}\,,\quad B = \Big(\frac{n-2}{(n-1)|L|^{2/n}}\Big)^{1/2} y
\end{equation}
from equation~(\ref{eq:gftogl}). It follows that $A^{-1}(B,B) = n-2 >  -1$; then the signature of $g^F$ is the signature of $A$, i.e., that of $g^L$ up to the sign of $L$. $\square$

\vspace{6pt}We summarize our construction of observers' frames and measurements into a precise definition. 

\vspace{6pt}\noindent\textbf{Definition 2.}
\textit{Let $(M,L,F)$ be a Finsler spacetime. Physical observers along worldlines $\tau\mapsto \gamma(\tau)$ in $M$ are described by a frame basis $\{e_\mu\}$ of $H_{(\gamma,\dot\gamma)}TM$ which
\begin{enumerate}[(i)]
\item has a timelike vector $e_0$ in the sense $\pi_*e_0 = \dot\gamma \in S_\gamma$; and
\item is $g^F$-orthogonal, $g^F_{(\gamma,\dot\gamma)}(e_\mu,e_\nu)=-\eta_{\mu\nu}$.
\end{enumerate}
They measure the components of horizontal tensor fields over $TM$ with respect to their frame at their tangent bundle position $(\gamma,\dot\gamma)$.}

\vspace{6pt}The Minkowski metric $\eta_{\mu\nu}$ in this definition has signature $(-1_1,1_3)$. We emphasize again that the frame $\{e_\mu\}$ in $H_{(\gamma,\dot\gamma)}TM$ can be identified one to one with a frame $\{\pi_* e_\mu\}$ in $T_\gamma M$, or reversely by the horizontal lift.  From~(\ref{eq:gftogl}), one can show that the observer frame has the nice property to diagonalize $g^F$ and $g^L$ simultaneously. We will now discuss the measurement procedure in more detail and present the example of how an observer measures the speed of light.

\subsection{Measuring the speed of light}\label{sec:measure}
Definition 2 of the observer frame includes the statement that a physical observable is given by the components of a horizontal tensor field with respect to the observer's frame, evaluated at her position on the tangent bundle, i.e., at her position on the manifold and her four-velocity. The motivation for this is as follows. The geometry of Finsler spacetimes is formulated on the tangent bundle $TM$, and hence matter tensor fields coupling to this gravitational background must also be defined over $TM$. Not all such tensor fields can be interpreted as tensor fields from the perspective of the spacetime manifold $M$. This interpretation requires that the tensor fields be horizontal; then they are multilinear maps built on the horizontal space $H_PTM$ and its dual which are identified with the tangent space $T_{\pi(P)}M$ and its dual. Consider the example of a 2-form field $\Phi$ over $TM$; in the horizontal/vertical basis this expands as
\begin{equation}\label{eq:2form}
\Phi = \Phi_{1\,ab}(x,y)\, dx^a \wedge dx^b +  2 \,\Phi_{2\,ab}(x,y)\, dx^a \wedge \delta y^b + \Phi_{3\,ab}(x,y)\, \delta y^a \wedge \delta y^b\,.
\end{equation} 
Only the purely horizontal part $\Phi_{1\,ab}(x,y)\, dx^a \wedge dx^b$ has a clear interpretation. Note that such horizontal tensor fields are automatically d-tensor fields, and have the same number of components as a tensor field of same rank on $M$. The difference is that the components depend on the tangent bundle position. The measurement of a horizontal tensor field by an observer at the tangent bundle position $(\gamma,\dot\gamma)$ clearly requires an observer frame of $H_{(\gamma,\dot\gamma)}TM$ in order to read out the components. 

We emphasize that the dependence of observables on the four-velocity of the observer is not surprising. Neither is it problematic as long as observers can communicate their results. In general relativity, observables are the components of tensor fields over $M$ with respect to the observer's frame in $T_\gamma M$; they clearly depend on $\dot\gamma$ which induces the splitting of $T_\gamma M$ into time and space directions. On Finsler spacetimes the dependence of observables on the observer's four-velocity is not only present in the time/space split of $H_{(\gamma,\dot\gamma)}TM$, but also in the argument of the tensor field components.

As a simple example we discuss the measurement of the spatial velocity of a point particle that moves on a worldline $\rho$ with horizontal tangent $\dot \rho$. This can be expanded in the orthonormal frame of an observer as $\dot\rho=\dot\rho^0 e_0+\dot{\vec\rho}=\dot\rho^0 e_0+\dot\rho^\alpha e_\alpha$, where we recall that $e_0=\dot\gamma$ is the observer's four velocity. The time $\dot\rho^0$ passes while the particle moves in spatial direction $\dot\rho^\alpha$, so the spatial velocity $\vec v$ and its square $v^2$ are
\begin{equation}\label{eq:v2}
\vec v=\frac{\dot{\vec\rho}}{\dot\rho^0}, \quad v^2=\frac{\delta_{\alpha\beta}\dot{\rho}^\alpha\dot{\rho}^\alpha}{(\dot\rho^0)^2}=-\frac{g^F_{(\gamma,\dot\gamma)}(\dot{\vec\rho},\dot{\vec\rho})}{g^F_{(\gamma,\dot\gamma)}(\dot\rho,\dot\gamma)^2}\,.
\end{equation}

As a consequence of this formula we may derive the speed of light seen by a given observer. As discussed in~\cite{Pfeifer:2011tk} light propagates on null worldlines $\rho$ with $L(\rho,\dot\rho)=0$ which is equivalent to $F(\rho,\dot\rho)^2=0$. We can use this fact to replace the Finsler metric in the formula for the velocity above. Taylor expanding $F(\rho,\dot\rho^0 e_0+\dot{\vec\rho})^2=0$ around $\dot{\vec \rho}=0$ yields
\begin{equation}\label{eq:light}
0=(\dot\rho^0)^2+ g^F_{(\rho,\dot\gamma)}(\dot{\vec\rho},\dot{\vec\rho})+\sum_{k=3}^\infty\frac{(\dot\rho^0)^{2-k}}{k!}\bar\partial_{c_1}...\bar\partial_{c_k}F(\rho, \dot\gamma)^2\dot{\vec\rho}^{c_1}...\dot{\vec\rho}^{c_k}.
\end{equation}
Evaluating this formula at the position of the observer $\rho=\gamma$ and dividing by $(\dot\rho^0)^2$, we immediately obtain an expression for the speed of light $c^2_{(\gamma, \dot\gamma)}(\dot{\vec\rho})$, i.e., the speed of light traveling in spatial direction $\dot{\vec\rho}$ and measured by the observer $(\gamma,\dot\gamma)$:
\begin{equation}\label{eq:speedlight}
c^2_{(\gamma, \dot\gamma)}(\dot{\vec\rho})=1+\sum_{k=3}^\infty\frac{(\dot\rho^0)^{-k}}{k!}\bar\partial_{c_1}...\bar\partial_{c_k}F(\gamma, \dot\gamma)^2\dot{\vec\rho}^{\ c_1}...\dot{\vec\rho}^{\ c_k}\,.
\end{equation}
The $\dot\rho^0$ are determined by solving the null condition $L(\gamma,\dot\rho^0\dot\gamma+ \dot{\vec\rho})=0$; on a generic Finsler spacetime there can be more than one solution since the null structure can be very complicated. From equation~(\ref{eq:speedlight}) we see that the measured speed of light depends on the higher than second order derivatives of the squared Finsler function; these vanish in the metric limit where we thus reobtain $c^2_{(\gamma, \dot\gamma)}(\dot{\vec\rho})=1$ independent of the observer and the spatial direction of the light ray.  The formulae~(\ref{eq:v2}) and~(\ref{eq:light}) enable us to compare experimental results on particle and light velocities with predictions on specific Finsler spacetime models. In~\cite{Pfeifer:2011ve} we used this to study the possibility of superluminal particle propagation.

\subsection{Generalized Lorentz transformations}
We already stressed the importance that observers should be able to communicate their measurements. Consider two observers whose worldlines meet at a point $x\in M$.  Since observers by Definition~2 measure the components of horizontal tensor fields in their frame and at their tangent bundle position, we need to determine which transformation uniquely maps an observer frame $\{e_\mu\}$ in $H_{(x,y)}TM$ to a second observer frame $\{f_\mu\}$ in $H_{(x,z)}TM$. Their respective four-velocities, or time directions, $y=\pi_*e_0$ and $z=\pi_*f_0$ generically are different, so that the  two observer frames are objects in tangent spaces to $TM$ at different points. As a consequence, we will now demonstrate that the transformations between observers consist of two parts: the first is a transport of the frame $\{e_\mu\}$ from $(x,y)$ to $(x,z)$, the second will turn out to be a Lorentz transformation.

\vspace{6pt}\noindent\textbf{Theorem~2.}
\textit{Consider two observer frames $\{e_\mu\}$ in $H_{(x,y)}TM$ and $\{f_\mu\}$ in $H_{(x,z)}TM$ on a Finsler spacetime $(M,L,F)$. If $z$ is in a sufficiently small neighbourhood around $y\in T_xM$, then the following procedure defines a unique map $\{e_\mu\}\mapsto\{f_\mu\}$:
\begin{enumerate}[(i)]
\item Let $t\mapsto v(t)$ be a vertical autoparallel of the Cartan linear connection  that connects ${v(0)=(x,y)}$ to $v(1)=(x,z)$; this satisfies $\pi_*\dot v=0$ and $\nabla_{\dot v}\dot v=0$. Determine a frame $\{\hat e_\mu(v(t))\}$ along $v(t)$ by parallel transport $\nabla_{\dot v} \hat e_\mu =0$ with the initial condition $\hat e_\mu(v(0))=e_\mu$.
\item Find the unique Lorentz transformation $\Lambda$ so that $f_\mu = \Lambda^\nu{}_\mu\hat e_\nu (v(1))$.
 \end{enumerate}}
 
\vspace{6pt}\noindent\textit{Proof.} We first show that the curve $v$ required in \textit{(i)} exists. The verticality condition $\pi_*\dot v=0$ implies $\dot v = \dot v^a\bar\partial_a$; the definition of the Cartan linear connection~(\ref{eq:cartanlin}) then tells us that $\nabla_{\dot v}\dot v=0$ is equivalent to solving $\ddot v^a + \frac{1}{2}g^{F\,ap}\bar\partial_p g^F_{bc}\dot v^b\dot v^c =0$. This has a unique solution connecting $(x,y)$ to any point $(x,z)$ in a sufficiently small neighbourhood in $T_xM$. Now let $\{\hat e_\mu(v(t))\}$ be the parallelly transported vector fields $\nabla_{\dot v}\hat e_\mu=0$ with $\hat e_\mu(v(0))=e_\mu$. The properties of the Cartan linear connection ensure that the $\hat e_\mu$ are horizontal fields. Observe also that $\nabla_{\dot v} \big(g^F_v(\hat e_\mu, \hat e_\nu)\big)=0$ along the curve $v$ since $g^F$ is covariantly constant under $\nabla$. It follows that
\begin{equation}
g^F_{v(t)}(\hat e_\mu(v(t)), \hat e_\nu(v(t)))=-\eta_{\mu\nu}
\end{equation}
is independent of $t$, and holds in particular at the final point of the transport $v(1)=(x,z)$. Now $\{\hat e_\mu(v(1))\}$ and $\{f_\mu\}$ are orthonormal frames with respect to $g^F$ in $H_{(x,z)}TM$; hence they are related by a unique Lorentz transformation as stated in point \textit{(ii)}~of the theorem. $\square$ 

\begin{figure}[h]
\vspace*{12pt}
\includegraphics[width=0.55\textwidth]{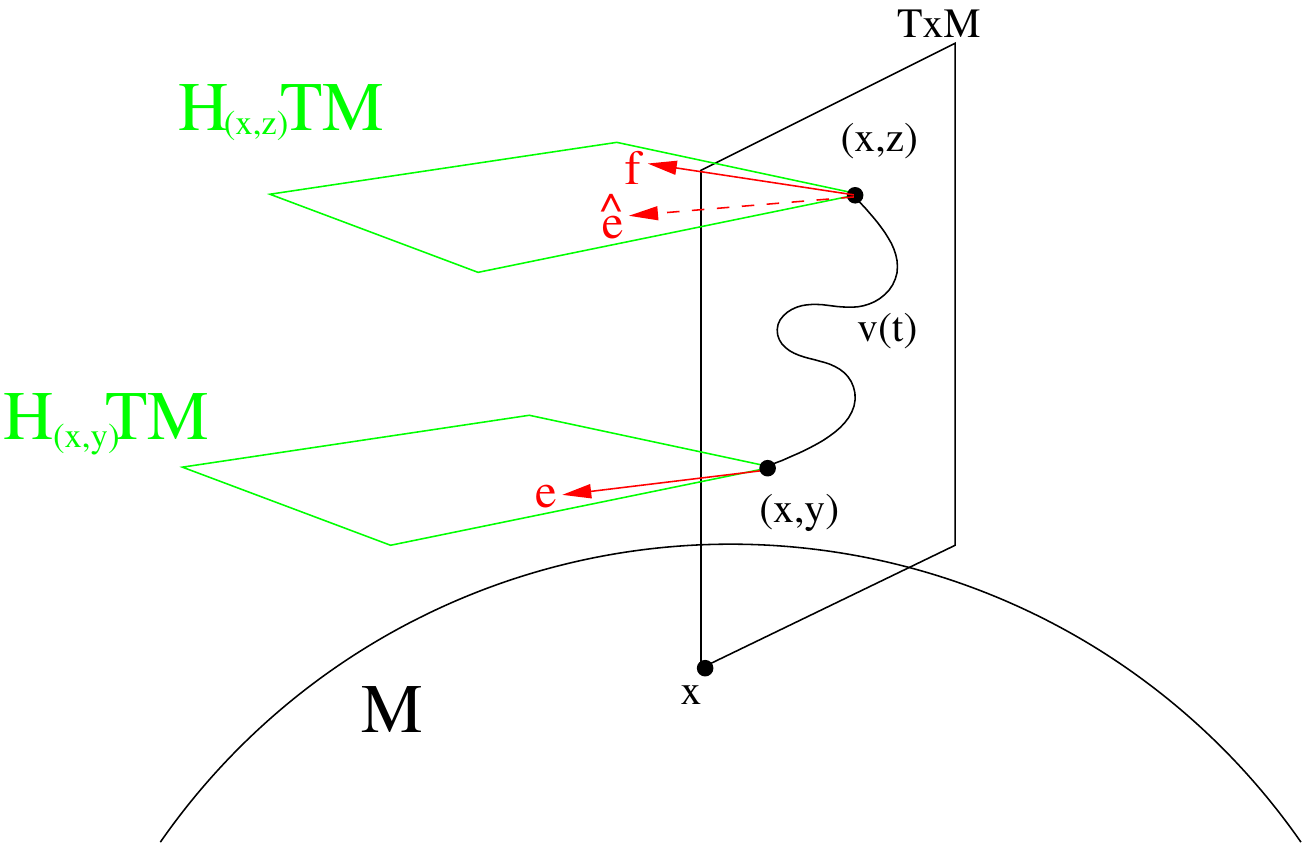}
\caption{\label{fig:trafo}\textit{Transformation between two observer frames: the frame $\{e_\mu\}$ in $H_{(x,y)}TM$ is first parallely transported to $\{\hat e_\mu\}$ in $H_{(x,z)}TM$, second Lorentz transformed into the final frame $\{f_\mu\}$.}}
\end{figure}

\vspace{6pt}The procedure described in Theorem~2 provides a map between the frames of two observers at the same point of the manifold $x\in M$, but with different four-velocities $y,z\in S_x\subset T_xM$; we display the two parts of this procedure as $\Lambda\circ P_{y\rightarrow z}$, i.e., as parallel transport followed by Lorentz transformation, which is illustrated in figure~\ref{fig:trafo}. The combined maps transform observers uniquely into one another as long as the autoparallel $v$ connecting the vertically different points in $TM$ exists and is unique. This is certainly the case if $(x,y)$ and $(x,z)$ are sufficiently close to each other. Whether the geometric structure of a specific, or maybe all, Finsler spacetimes is such that unique transformations between all observers exist requires requires further investigation. 

In the observer transformations on generic Finsler spacetimes there appears an additional ingredient that is not present on metric spacetimes. Before applying the Lorentz transformation to the frame, one has to perform a parallel transport in the vertical tangent space. In the metric limit the vertical covariant derivative becomes trivial so that the parallely transported frame does not change at all along the curve $v$. In this special case the transformation of an observer thus reduces to $\Lambda\circ \textrm{id}_{y\rightarrow z}$ which is fully determined by a Lorentz transformation.

The observer transformations on Finsler spacetimes essentially have the algebraic structure of a groupoid that reduces to the Lorentz group in the metric limit. We first review the general definition of a groupoid and then show how this applies to our case.

\vspace{6pt}\noindent\textbf{Definition 3.}
\textit{A groupoid $\mathcal{G}$ consists of a set of objects $G_0$ and a set of arrows $G_1$. Every arrow $A$ is assigned a source $e=s(A)$ and a target $f=t(A)$ by the maps $s:G_1\rightarrow G_0$ and $t:G_1\rightarrow G_0$; one writes this as $A:e\rightarrow f$. For arrows $A$ and $B$ whose source and target match as $t(A)=s(B)$ there exists an associative multiplication $G_1\times G_1\rightarrow G_1, (A,B)\mapsto BA$ with
\begin{equation}\label{eq:mult}
s(BA)=s(A)\,,\quad t(BA)=t(B)\,,\quad C(BA) = (CB)A\,.
\end{equation}
A unit map $G_0\rightarrow G_1,e\mapsto \openone_e$  where $\openone_e:e\rightarrow e$ exists so that
\begin{equation}\label{eq:one}
\openone_{t(A)}A = A = A \,\openone_{s(A)}\,.
\end{equation}
For every arrow $A$ exists an inverse arrow $A^{-1}$ that satisfies
\begin{equation}\label{eq:inv}
s(A^{-1})=t(A)\,,\quad t(A^{-1})=s(A)\,,\quad 
A^{-1}A = \openone_{s(A)}\,, \quad A A^{-1} =\openone_{t(A)}\,.
\end{equation}}
 
\noindent Groupoids are generalizations of groups. These can be expressed as groupoids with a single object in $G_0$; then the arrows correspond to group elements all of which can be multiplied since sources and targets always match. The multiplication is associative, the identity element and inverse elements exist.

Consider $G_0=S_x\subset T_xM$ as the set of unit timelike vectors which contains the different four-velocities of observers at the point $x\in M$. Let the arrows in $G_1$ be the set of all maps between two observer frames at $x$ which are defined by the procedure stated in Theorem~2. In case the involved vertical autoparallels connect the four-velocities uniquely, the sets $G_0$ and $G_1$ define a groupoid: source and target of a map $A=\Lambda\circ P_{y\rightarrow z}$ between two frames $\{e_\mu\in H_{(x,y)}TM\}$ and $\{f_\mu \in H_{(x,z)}TM\}$ are simply given by $s(A)=y\in S_x$ and $t(A)=z\in S_x$; the multiplication $BA$ is defined by applying the procedure of Theorem~2 to construct the map between $s(A)$ and $t(B)$, which gives the properties~(\ref{eq:mult}); we choose the unit map $\openone_{y}$ that provides~(\ref{eq:one}) as $\openone_{y} = \openone\circ \textrm{id}_{y\rightarrow y}$, i.e., as trivial parallel transport of the frame $\{e_\mu\in H_{(x,y)}TM\}$ with respect to the Cartan linear connection along the vertical autoparallel that stays at $(x,y)$ followed by the identity Lorentz transformation. Finally, we define the inverse $A^{-1}= \Lambda^{-1}\circ P_{z\rightarrow y}$, where $P_{z\rightarrow y}$ denotes parallel transport backwards along the unique vertical autoparallel connecting $(x,y)$ and $(x,z)$ which is also used for $P_{y\rightarrow z}$; to check the properties~(\ref{eq:inv}), one simply shows that parallel transport of the frames and Lorentz transformation commute. Thus  we have shown the following result:

\vspace{6pt}\noindent\textbf{Theorem 3.}
\textit{On Finsler spacetimes $(M,L,F)$ the transformations between observer frames ${\{e_\mu\in H_{(x,y)}TM\}}$ at $x\in M$ that are attached to points $(x,y)\in U_x\subset S_x$ define a groupoid $\mathcal{G}$ under the condition that any pair of points in $U_x$ can be connected by a unique vertical autoparallel of the Cartan linear connection.}

\vspace{6pt}We already discussed that the transformations of observer frames reduce to the form $A=\Lambda\circ \textrm{id}_{y \rightarrow z}$ in the limit of metric geometry.  Hence the only information contained in the reduced groupoid $\tilde{\mathcal{G}}$ with $\tilde G_0=S_x$ and $\tilde G_1=\{\Lambda\circ \textrm{id}_{y\rightarrow z}\}$ is given by the Lorentz transformations. In mathematically precise language this can be expressed as the equivalence of $\tilde{\mathcal{G}}$ to the Lorentz group seen as a groupoid $\mathcal{H}$ with a single object $H_0=\{x\}$ and arrows $H_1=\{\Lambda\}$. The functor $\varphi:\tilde{\mathcal{G}}\rightarrow \mathcal{H}$ establishing the equivalence can be defined by the projection $\varphi_0=\pi:\tilde G_0\rightarrow H_0$ and by $\varphi_1:\tilde G_1\rightarrow H_1, \Lambda\circ \textrm{id}_{y\rightarrow z}\mapsto \Lambda$. Indeed, $\varphi$ can be checked to be injective, full and essentially surjective, and so it makes $\tilde {\mathcal{G}}$ and $\mathcal{H}$ equivalent. See~\cite{Baez:2008hx} for details on the required mathematical definitions.

\section{Gravitational dynamics}\label{sec:gravdyn}
We have now reviewed the basic concept of Finsler spacetime geometry and laid the foundations for the interpretation of physics on these backgrounds. We have seen that well-defined observers exist which communicate with each other by means of groupoid transformations that generalize the Lorentz group in metric geometry. As emphasized in previous work~\cite{Pfeifer:2011tk,Pfeifer:2011ve}, physical predictions in our generalized geometric framework require gravitational dynamics for the fundamental geometry function $L$ to determine specific spacetime solutions. In this section we for the first time present a Finsler gravity action along with a consistent minimal coupling principle between gravity and matter. We begin our presentation with the variation of the pure gravity action, before the coupling of Finsler gravity to matter is discussed in section~\ref{sec:gravmat}; the full field equation is derived in section~\ref{sec:gravmateq}.
Moreover, we prove that the Finsler gravity field equation becomes equivalent to the Einstein equations in the metric limit. 

\subsection{Action and vacuum equations}\label{sec:gravvac}
On Finsler spacetime the simplest curvature scalar built from the non-linear curvature tensor $R^a{}_{bc}$, which is relevant for the tidal acceleration of Finsler geodesics, is $\mathcal{R}=R^a{}_{ab}y^b$. This contains the lowest number of derivatives on the fundamental function~$L$ without involving additional d-tensors besides the curvature, like $S^a{}_{bc}$ or $\bar \partial_a g^F_{bc}$. Recall that integrals are well-defined over the seven-dimensional unit tangent bundle $\Sigma$ with coordinates $(\hat x,u)$, as discussed in section~\ref{sec:int}. These two facts directly lead to our Finsler gravity action
\begin{equation}\label{eq:gravact}
 S_G[L]=\int d^4\hat xd^3u \left[\sqrt{g^Fh^F}\ \mathcal{R}\right]_{|\Sigma}\,.
\end{equation}

The gravitational field equation in vacuum now is obtained by variation with respect to $L$. To perform this variation for an $m$-homogeneous function $f(x,y)$ on $TM$ restricted to $\Sigma$ it is useful to realise that
\begin{equation}
\delta(f_{|\Sigma})=(\delta f)_{|\Sigma}-\frac{m}{n}f_{|\Sigma}\frac{\delta L}{L}\,,
\end{equation}
where $n$ is the homogeneity of $L$.
With the help of this formula and the results for integration by parts in~(\ref{eq:intbp}) we can derive the vacuum field equations in three steps. The first uses the variation formula above with $f(x,y)=\sqrt{g^Fh^F}\ R^a{}_{ab}y^b$ and $m=5$, which yields
\begin{eqnarray}\label{eq:calcgrav1}
\delta S_G[L]=\int d^4\hat xd^3u \left[\delta\Big(\sqrt{g^Fh^F}\ \mathcal{R}\Big)-\frac{5}{n}\sqrt{g^Fh^F}\ \mathcal{R} \frac{\delta L}{L}\right]_{|\Sigma}\,.
\end{eqnarray}
The second step is the variation of the volume element which leads to
\begin{equation}\label{eq:calcgrav2}
\delta S_G[L]=\int d^4\hat xd^3u \sqrt{g^Fh^F}_{|\Sigma} \left[ \Big(g^{F\,pq}\delta g^F_{pq}-\frac{6}{n}\frac{\delta L}{L}\Big)\mathcal{R}+y^b\delta R^a{}_{ab} \right]_{|\Sigma}\,,
\end{equation}
while in the third step we use the following identities
\begin{subequations}\label{eq:calcgrav}
\begin{equation}\label{eq:calcgrav31}
\int d^4\hat xd^3u \left[\sqrt{g^Fh^F}\ g^{F\,pq}\delta g^F_{pq}\mathcal{R}\right]_{|\Sigma} = \int d^4\hat xd^3u \left[\sqrt{g^Fh^F}\ g^{F\,ab}\bar\partial_a\bar\partial_b\mathcal{R}\frac{\delta L}{nL}\right]_{|\Sigma}\,,
\end{equation}
\begin{equation}\label{eq:calcgrav32}
\int d^4\hat xd^3u \left[\sqrt{g^Fh^F}\ y^b\delta R^a{}_{ab}\right]_{|\Sigma} = \int d^4\hat xd^3u \left[\sqrt{g^Fh^F}\ 2g^{F\,ab}\big(\nabla_aS_b+S_aS_b+\bar\partial_a\nabla S_b\big)\frac{\delta L}{nL}\right]_{|\Sigma}\,,
\end{equation}
\end{subequations}
to arrive at the final form of the variation of the Finsler gravity action~(\ref{eq:gravact}):
\begin{equation}
\delta S_G[L]=\int d^4\hat xd^3u \sqrt{g^Fh^F}_{|\Sigma}\left[ g^{F\,ab}\bar\partial_a\bar\partial_b\mathcal{R}-6\mathcal{R}+2g^{F\,ab}\big(\nabla_aS_b+S_aS_b+\bar\partial_a\nabla S_b\big)\right]_{|\Sigma}\frac{\delta L}{nL}\,.
\end{equation}
For further details of this variation we refer the reader to appendix~\ref{app:vargrav}. Now we can read off the vacuum field equation on $\Sigma$ as
\begin{equation}
\left[ g^{F\,ab}\bar\partial_a\bar\partial_b\mathcal{R}-6\mathcal{R}+2g^{F\,ab}\big(\nabla_aS_b+S_aS_b+\bar\partial_a\nabla S_b\big)\right]_{|\Sigma} = 0 \,.
\end{equation}

Observe that all terms in the bracket are zero-homogeneous on $TM$, except the second term~$\mathcal{R}$ that has homogeneity two. Since $(\mathcal{R})_{|\Sigma}= (\mathcal{R}/F^2)_{|\Sigma}$ we can replace the second term by $\mathcal{R}/F^2$ which is now also zero-homogeneous. Hence the equation can be lifted to $TM$ in the form
\begin{equation}\label{eq:gravvac}
g^{F\,ab}\bar\partial_a\bar\partial_b\mathcal{R}-\frac{6}{F^2}\mathcal{R}+2g^{F\,ab}\big(\nabla_aS_b+S_aS_b+\bar\partial_a\nabla S_b\big)=0\,.
\end{equation}
It seems as if this equation could be invalid on $\{L=0\}=\{F=0\}$ where $F$ is not differentiable so that the Finsler metric $g^F$ does not exist. However, this is not the case: the equation is valid also on the null structure. To see this, one expresses $g^F$ through $g^L$ with the help of formula~(\ref{eq:gftogl}) and multiplies by $F^2$. The resulting equation is well-defined whereever $g^L$ is nondegenerate, and in particular on the null structure. Note that equation~(\ref{eq:gravvac}) is invariant under the transformation $L \rightarrow L^k$, which will be a guiding principle for  matter coupling below.

In the metric limit $L=g_{ab}(x)y^ay^b$, the tensors in the the Finsler gravity equation reduce as $\mathcal{R}=-y^ay^b R_{ab}$ and $S_a=0$, where $R_{ab}$ is the Ricci tensor of the metric $g$. Accordingly, the field equation becomes
\begin{equation}
2 R+\frac{6}{F^2}R_{ab}y^ay^b=0
\end{equation} 
which is equivalent to the Einstein vacuum equations $R_{ab}=0$ by differentiating twice with respect to $y$. We conclude that a family of solutions of the Finsler gravity vacuum equation~(\ref{eq:gravvac}) is induced by solutions
 $g_{ab}(x)$ of the vacuum Einstein equations via the fundamental functions ${L_k=(g_{ab}(x)y^ay^b)^k}$. In section~\ref{sec:Corr} we will present a solution of the Finsler gravity vacuum equation beyond metric geometry.

\subsection{Consistent matter coupling}\label{sec:gravmat}
Above we have achieved a consistent generalization of vacuum Einstein gravity from metric spacetimes to Finsler spacetimes. Next we will show that this generalization can be completed by the coupling of matter fields. For this purpose we will discuss a minimal coupling principle that generates consistent matter field actions on Finsler spacetimes from their well-known counterparts on metric spacetimes. In the discussion we restrict our attention to $p$-form fields; spinor fields have to be investigated further. In section~\ref{sec:gravmateq} we will deduce the complete gravity equations with energy-momentum source term.

Consider an action $S_m[g,\phi]$ for a physical $p$-form field $\phi$ on a Lorentzian spacetime $(M,g)$,
\begin{equation}
\tilde S_m[g,\phi] =\int_M d^4x\ \sqrt{g}\ \mathcal{L}(g,\phi, \mathrm{d}\phi)\,.
\end{equation}
The corresponding matter action on Finsler spacetime is obtained by lifting $\tilde S_m$ to the tangent bundle $TM$ equipped with the Sasaki-type metric $G(x,y)$ defined in~(\ref{eq:sasaki}) in the following way:
\begin{enumerate}[(i)]
 \item consider the Lagrangian density $\mathcal{L(\dots)}$ of the standard theory on $M$ as a contraction prescription that forms a scalar function from various tensorial objects;
 \item replace the Lorentzian metric $g(x)$ in $\mathcal{L(\dots)}$ by the Sasaki-type metric $G(x,y)$; 
 \item replace the $p$-form field $\phi(x)$ on $M$ by a zero-homogeneous $p$-form field\footnote{A two-form field $\Phi$ as in~(\ref{eq:2form}), for example, is zero-homogeneous if and only if its components $\Phi_1$, $\Phi_2$ and $\Phi_3$ have the homogeneities $0$, $-1$ and $-2$, respectively; these are cancelled by the homogeneity of $\delta y$.} $\Phi(x,y)$ on $TM$;
 \item introduce Lagrange multipliers $\lambda$ for all not purely horizontal components of $\Phi$;
 \item finally integrate over the unit tangent bundle $\Sigma$ with the volume form given by the pull-back~$G^*$ of the Sasaki-type metric. 
\end{enumerate}
The result of this procedure is the Finsler spacetime field theory action
\begin{equation}\label{eq:matact} 
S_m[L,\Phi, \lambda]=\int_\Sigma d^4\hat xd^3u \Big[\sqrt{g^Fh^F} \Big(\mathcal{L}(G,\Phi, \mathrm{d}\Phi)+\lambda (1-P^H)\Phi \Big)\Big]_{|\Sigma}\,.
\end{equation}

The projection $P^H$ projects to the purely horizontal part of the $p$-form $\Phi$; in the example of the general two-form on $TM$ displayed in equation~(\ref{eq:2form}) we have 
\begin{equation} 
P^H\Phi = \Phi_{1\,ab}\,dx^a\wedge dx^b\,.
\end{equation}
The Lagrange multiplier guarantees that the on-shell degrees of freedom of $\Phi$ are precisely those with a clear physical interpretation as fields along the manifold $M$, as discussed in section~\ref{sec:measure}. The minimal coupling principle for matter to Finsler spacetime presented above is a  slightly refined version of  that in~\cite{Pfeifer:2011tk} where we discussed electrodynamics on Finsler spacetime. The only modification here is in the definition of the Sasaki-type metric $G(x,y)$ that is now defined in terms of the Finsler metric $g^F$ instead of~$g^L$.  This change ensures that the resulting matter action $S_m[L,\Phi,\lambda]$ is invariant under $L\rightarrow L^k$ in the same way as  the pure gravity action. Nevertheless, the results obtained for electrodynamics on Finsler spacetime in~\cite{Pfeifer:2011tk} are  unchanged by the refined coupling principle presented here.

The matter field equations obtained by extremizing the action with respect to the $p$-form field~$\Phi$ and the Lagrange multiplier $\lambda$ can be studied most easily if expressed in components with respect to the horizontal/vertical basis. The calculation is performed in detail in appendix~\ref{app:varmat}. We display the results with the convention that barred indices denote vertical components, unbarred indices now denote horizontal components, and capital indices both horizontal and vertical components. Variation with respect to the Lagrange multiplier yields the constraints
\begin{equation}\label{eq:cons}
\Phi_{\bar a_1...\bar a_i a_{i+1}...a_p} = 0\,, \quad\forall i=1\dots p\,. 
\end{equation}
Variation for the purely horizontal components of $\Phi$ gives
\begin{equation}\label{eq:eom1}
\frac{\partial \mathcal{L}}{\partial \Phi_{a_1...a_p}}-(p+1)(\nabla_q +S_q)\frac{\partial \mathcal{L}}{\partial (\mathrm{d}\Phi_{qa_1...a_p})}-(\bar\partial_{\bar q}+g^{F\,mn}\bar\partial_{\bar q}g^F_{mn}-4g^F_{\bar{q}q}y^q) \frac{\partial \mathcal{L}}{\partial (\mathrm{d}\Phi_{\bar{q}a_1...a_p})}=0
\end{equation}
which determines the evolution of the physical field components, while variation with respect to the remaining components produces
\begin{eqnarray}
\lambda^{\bar a_1A_2...A_p} &=& - \frac{\partial \mathcal{L}}{\partial \Phi_{\bar a_1A_2...A_p}} 
+ (p+1)(\nabla_q +S_q)\frac{\partial \mathcal{L}}{\partial (\mathrm{d}\Phi_{q\bar a_1A_2...A_p})}
+ \frac{p(p+1)}{2} \frac{\partial \mathcal{L}}{\partial (\mathrm{d}\Phi_{PQ A_2...A_p})}\gamma^{\bar a_1}{}_{PQ}\nonumber \\
&&+ (\bar\partial_{\bar q}+g^{F\,mn}\bar\partial_{\bar q}g^F_{mn}-4g^F_{\bar{q}q}y^q) \frac{\partial \mathcal{L}}{\partial (\mathrm{d}\Phi_{\bar{q}\bar{a}_1A_2...A_p})}\label{eq:eom2}
\end{eqnarray}
which fixes the components of the Lagrange multiplier. The $\gamma^{\bar a}{}_{PQ}$ are the commutator coefficients of the horizontal/vertical basis. 

Our coupling principle is consistent with the metric limit, i.e., the equations of motion obtained from the Finsler spacetime action reduce to the equations of motion on Lorentzian spacetime in the case $L=g_{ab}(x)y^ay^b$  and $\Phi_{A_1...A_p}(x,y)=\phi_{A_1...A_p}(x)$. Then we have the geometric identity $S_a=0$; moreover 
\begin{equation}
\mathrm{d}\Phi_{a_1...a_{p+1}}=(p+1)\partial_{[a_1}\phi_{a_2...a_{p+1}]} 
\end{equation}
using the constraints~(\ref{eq:cons}) and the fact that the horizontal derivative acts as a partial derivative on the $y$-independent $p$-form components. Finally,
\begin{equation}
\frac{\partial \mathcal{L}}{\partial (\mathrm{d}\Phi_{\bar{q}a_1...a_p})}=0
\end{equation}
because, as a consequence of our coupling principle where the Sasaki-type metric is block-diagonal in the horizontal/vertical basis, the vertical index of $\mathrm{d}\Phi_{\bar{q}a_1...a_p}$ must appear in $\mathcal{L}(G,\Phi, \mathrm{d}\Phi)$ contracted via $g^F$ into either a vertical derivative or into components of $\Phi$ with at least one vertical index. In the metric limit, vertical derivatives give zero, while the constraints~(\ref{eq:cons}) guarantee that all components of $\Phi$ with at least one vertical index vanish. Combining these observations shows that equation~(\ref{eq:eom1}) reduces to
\begin{equation}\label{eq:red}
\frac{\partial \mathcal{L}}{\partial \Phi_{a_1...a_p}}-(p+1)\nabla_q\frac{\partial \mathcal{L}}{\partial (\mathrm{d}\Phi_{qa_1...a_p})} = 0\,,
\end{equation}
where $\nabla$ now operates in the same way as the Levi--Civita connection of the metric $g$. Again, as a consequence of our minimal coupling principle with the block-diagonal form of the Sasaki-type metric in the horizontal/vertical basis, we can conclude in the metric limit that 
\begin{equation}
\frac{\partial \mathcal{L}(G,\Phi,\mathrm{d}\Phi)}{\partial \Phi_{a_1...a_p}} = \frac{\partial \mathcal{L}(g,\phi,\mathrm{d}\phi)}{\partial \phi_{a_1...a_p}}\,,\quad \frac{\partial \mathcal{L}(G,\Phi,\mathrm{d}\Phi)}{\partial (\mathrm{d}\Phi_{qa_1...a_p})} = \frac{\partial \mathcal{L}(g,\phi,\mathrm{d}\phi)}{\partial (\mathrm{d}\phi_{qa_1...a_p})} 
\end{equation}
so that (\ref{eq:red}) becomes equivalent to the standard $p$-form field equation of motion on metric spacetime. 

Our minimal coupling procedure for matter fields to Finsler spacetime can be applied immediately for instance to the scalar field, as done in~\cite{Pfeifer:2011ve}. Note that it can be easily extended to the case of interacting form fields of any degree with metric spacetime action
\begin{equation}
\tilde S_m[g,\phi_1,\phi_2, ...] =\int_M d^4x\ \sqrt{g}\ \mathcal{L}(g,\phi_1, \mathrm{d}\phi_1,\phi_2,\mathrm{d}\phi_2,...)\,.
\end{equation}
The minimal coupling procedure then leads to the action
\begin{eqnarray}
S_m[L,\Phi_1, \lambda_1,&\Phi_2&,\lambda_2,...]\nonumber\\
&=& \int_\Sigma d^4\hat xd^3u \Big[\sqrt{g^Fh^F} \Big(\mathcal{L}(G,\Phi_1, \mathrm{d}\Phi_1,\Phi_2, \mathrm{d}\Phi_2,...)+\sum_I \lambda_I (1-P^H)\Phi_I \Big)\Big]_{|\Sigma}\,.
\end{eqnarray}
The equations of motion for each field $\phi_I$ have the same form as in the single field case, and the metric limit leads to the standard field equations by arguments that proceed in a completely analogous way as before.

In the standard formulation of electrodynamics, the action is a functional $\tilde S_m[g,A,dA]$ of a one-form potential $A$, but the classical physical field is $F=dA$. Our minimal coupling principle to obtain an action on Finsler spacetimes cannot be applied immediately to this situation: the problem is that the Lagrange multiplier then only kills the vertical components of the lift of $A$, but does not guarantee that the lift of $F$ is purely horizontal. This problem is solved in~\cite{Pfeifer:2011tk} by starting from an equivalent interacting action of the form $\tilde S_m[g,A,dA,F,dF]$ which provides the complete set of Maxwell equations $F=dA$ and $d\star_g F=0$ by variation. Now the minimal coupling principle entails that both the lifted fields $A$ and $F$ are purely horizontal and can be interpreted physically. 

\subsection{Gravity field equations and metric limit}\label{sec:gravmateq}
We are now in the position to study the interplay between the matter actions $S_m$ introduced in~(\ref{eq:matact}) and the pure Finsler gravity action $S_G$ in~(\ref{eq:gravact}). Their sum provides a complete description of gravity and classical matter fields on Finsler spacetimes:
\begin{eqnarray}\label{eq:compact}
S[L,\Phi,\lambda] & =& \kappa ^{-1}S_G[L]+S_m[L,\Phi,\lambda] \nonumber\\
& =& \kappa^{-1}\int d^4\hat xd^3u \left[\sqrt{g^Fh^F}\ \mathcal{R}\right]_{|\Sigma}+ \int d^4\hat xd^3u\ \Big[\sqrt{g^Fh^F}\ (\mathcal{L}(G,\Phi, \mathrm{d}\Phi)+\lambda(1-P^H)\Phi)\Big]_{|\Sigma}\,.
\end{eqnarray}
As usual, the matter field equations following from this are the same as for the pure matter action. The gravitational field equations are obtained by variation with respect to the fundamental geometry function~$L$. The variation of $S_m$ with respect to~$L$ is
\begin{equation}
\delta S_m  =  \int d^4\hat xd^3u\, \Big(\frac{\delta S_m}{\delta L}\delta L\Big)_{|\Sigma} = \int d^4\hat xd^3u\, \bigg(\sqrt{g^Fh^F}\frac{nL}{\sqrt{g^Fh^F}}\frac{\delta S_m}{\delta L}\bigg)_{|\Sigma} \frac{\delta L}{nL}\,, 
\end{equation}
and leads us to the definition of the energy momentum scalar $T_{|\Sigma}$ on the unit tangent bundle as
\begin{equation}\label{eq:EMS}
 T_{|\Sigma}=\bigg(\frac{nL}{\sqrt{g^Fh^F}}\frac{\delta S_m}{\delta L}\bigg)_{|\Sigma}\,.
\end{equation}
With this definition the complete gravitational field equations on Finsler spacetime including energy-momentum sources formally become
\begin{equation}\label{eq:fgrav}
\Big[g^{F\,ab}\bar\partial_a\bar\partial_b\mathcal{R}-\frac{6}{F^2}\mathcal{R}+2g^{F\,ab}\big(\nabla_aS_b+S_aS_b+\bar\partial_a\nabla S_b\big)\Big]_{|\Sigma}=-\kappa T_{|\Sigma}\,.
\end{equation}
As in the vacuum case with $T_{|\Sigma}=0$, these equations can be lifted to $TM$. The terms in the bracket on the left hand side are all zero-homogeneous and can be lifted trivially. The terms in $T$ without the restriction on the right hand side in principle can result from variation with different homogeneities; to lift these one simply multiplies each term by the appropriate power of $F$ in order to make it zero homogeneous. This is the same procedure applied in section~\ref{sec:gravvac} to the gravity side. 

The gravitational constant $\kappa$ will now be determined so that the gravitational field equation on Finsler spacetimes becomes equivalent to the Einstein equations in the metric limit. Variation with respect to $L$ of the concrete form of the matter action in~(\ref{eq:compact}) and performing the metric limit, i.e., $g^F{}_{ab}(x,y)=-g_{ab}(x)$ for observers and $\Phi_{A_1 ... A_p}(x,y)=\phi_{A_1 ... A_p}(x)$, the gravity equation~(\ref{eq:fgrav})  becomes
\begin{equation}\label{eq:tsig}
2g^{ab}R_{ab}+6\frac{R_{ab}y^ay^b}{|g_{pq}y^py^q|}=-\kappa\Big( 4\mathcal{L}-4g_{ab}\frac{\partial \mathcal{L}}{\partial g^{}_{ab}}-24\frac{ y_ay_b}{|g_{pq}y^py^q|}\frac{\partial \mathcal{L}}{\partial g^{}_{ab}}\Big)\,.
\end{equation}
The detailed calculation of this result is involved and can be found in appendix~\ref{app:varmat}. Introducing the standard energy momentum tensor of $p$-form fields on Lorentzian metric spacetimes $\tilde T^{ab}=g^{ab}\mathcal{L}+2\frac{\partial \mathcal{L}}{\partial g_{ab}}$ and its trace $\tilde T=\tilde T^{ab} g_{ab}=4\mathcal{L}+ 2 g_{ab}\frac{\partial \mathcal{L}}{\partial g_{ab}}$ we can rewrite the equation above as 
\begin{equation}\label{eq:gravlimit}
2 R-6\frac{R_{ab}y^ay^b}{g_{pq}y^py^q}=-\kappa\Big(-2 \tilde T+12 \frac{\tilde T^{ab}y_ay_b}{g_{pq}y^py^q}\Big),
\end{equation}
if evaluated at $g$-timelike observer four-velocities $y$. Now we take a second derivative with respect to $y$, contract with $g^{-1}$, reinsert the result, and conclude
\begin{equation}
\Big(R_{ab}-\frac{1}{2}g_{ab}R\Big)y^ay^b= 2\kappa\ \tilde T_{ab}y^ay^b\,.
\end{equation}
Since there is no $y$-dependence beyond the explicit one, a second derivative with respect to $y$ yields the Einstein equations, if we choose  the gravitational constant $\kappa=\frac{4\pi G}{c^4}$. 

The gravity equation on Finsler spacetime including the coupling to matter therefore is
\begin{equation}\label{eq:complGrav}
-g^{Fab}\bar\partial_a\bar\partial_b \mathcal{R}+\frac{6}{F^2}\mathcal{R}-2g^{F\,ab}\big(\nabla_aS_b+S_aS_b+\bar\partial_a\nabla S_b\big)=\frac{4\pi G}{c^4} T
\end{equation}
with zero homogeneous source function $T$ on $TM$. Observe that this field equation including the matter part is invariant under $L \rightarrow L^k$ by construction of the coupling principle, as the vacuum equation is. This leads to the interesting conclusion that every solution $g_{ab}(x)$ of the Einstein equations induces a family $L_k$ of solutions of the Finsler gravity solution with $L_k=(g_{ab}(x)y^ay^b)^k$.

In order to find further solutions of this highly complicated partial differential equation we will study symmetries of Finsler spacetimes in the next section. Then we present a solution of the linearised Finsler gravity equation in section~\ref{sec:Corr} which turns out to be a geometric refinement of the linearised Schwarzschild solution of general relativity.

\section{Finsler spacetime symmetries}\label{sec:fsym}
In the previous section we have deduced the Finsler gravity field equation including matter sources and shown that it is consistent with the Einstein equations in the metric geometry limit. Our new field equation is a highly complex differential equation; in order to simplify the task of finding analytic solutions we wish to consider symmetric spacetimes. We begin this section by defining symmetries of Finsler spacetimes and show how this concept is a generalisation of the symmetries of Lorentzian spacetimes. We explicitly present the general structure of the fundamental geometry function $L$ for the spherically, cosmologically and maximally symmetric case. In the next section we will then use our results about symmetric Finsler spacetimes to solve the linearised Finsler gravity equation.

\subsection{Definition}
On a symmetric Lorentzian manifold $(M,g)$, the metric is invariant under certain diffeomorphisms; similarly we wish to define symmetries of a Finsler spacetime $(M,L,F)$ as an invariance of the fundamental geometry function~$L$.  Consider a diffeomorphism generated by the vector field $X=\xi^a(x)\partial_a$; this acts as a coordinate change on local coordinates on $M$ as $(x^a)\rightarrow(x^a+\xi^a)$, and on the induced coordinates on the tangent bundle $TM$ as $(x^a,y^a)\rightarrow(x^a+\xi^a,y^a+y^q\partial_q\xi^a)$. Hence the diffeomophism on $M$ induces a diffeomorphism on $TM$ that is generated by the vector field $X^C=\xi^a\partial_a+y^q\partial\xi^a\bar\partial_a$, called the complete lift of $X$. The idea of implementing symmetries via complete lifts in a Finsler geometry setting appears already in~\cite{Li:2010wv}; here we want to make this concept precise for Finsler spacetimes:

\vspace{6pt}\noindent\textbf{Definition 4.}
\textit{A symmetry of a Finsler spacetime $(M,L,F)$ is a diffeomorphism generated by a vector field $Y$ over the tangent bundle $TM$ so that $Y(L)=0$ and $Y$ is the complete lift $X^C$ of a vector field $X$ over $M$. A Finsler spacetime is called symmetric if it possesses at least one symmetry.}

\vspace{6pt}The following theorem summarizes important properties of Finsler spacetime symmetries. The symmetry generators form a Lie algebra with the commutator of vector fields on $TM$, and they are isomorphic to a Lie algebra of vector fields on $M$ which becomes the usual symmetry algebra of Lorentzian manifolds in the metric geometry limit. This not only shows that Definition~4 of symmetry is consistent with that of Lorentzian spacetimes, but also that the usual Killing vectors, e.g., those for spherical symmetry, can be used to study symmetries of Finsler spacetimes. 

\vspace{6pt}\noindent\textbf{Theorem 4.} 
\textit{Let $\mathcal{S}$ be the set of symmetry-generating vector fields of a Finsler spacetime. 
\begin{enumerate}[(i)]
\item $(\mathcal{S},[\cdot,\cdot])$ is a Lie subalgebra of the set of vector fields over $TM$;
\item $(\mathcal{S},[\cdot,\cdot])$ is isomorphic to the Lie subalgebra $(\pi_*(\mathcal{S}),[\cdot,\cdot])$ of the set of vector fields over $M$;
\item in the metric geometry limit, $(\pi_*(\mathcal{S}),[\cdot,\cdot])$ becomes the symmetry algebra of the emerging Lorentzian spacetime. 
\end{enumerate}}

\noindent\textit{Proof.} {\it (i)} Let $Y\in\mathcal{S}$; then $Y(L)=0$ and $(\pi_*Y)^C-Y=0$. Both properties are linear, so that $\mathcal{S}$ is a vector subspace of the Lie algebra of all vector fields on $TM$. It remains to be proven that the commutator of two elements $Y_1,Y_2\in\mathcal{S}$ closes in $\mathcal{S}$. It is clear that $[Y_1,Y_2](L)=0$; to show that $(\pi_*[Y_1,Y_2])^C=[Y_1,Y_2]$, one uses that $Y_i=X_i^C$ for some vector fields $X_i$ on $M$ and that $[X_1^C,X_2^C]=[X_1,X_2]^C$, see~\cite{YanoIshihara}. 

{\it (ii)} The inverse for $\pi_*$ on $\pi_*(\mathcal{S})$ is given by the complete lift, hence $\mathcal{S}$ and $\pi_*(\mathcal{S})$ are isomorphic as vector spaces. The Lie algebra structure is preserved in both directions because of $[X_1^C,X_2^C]=[X_1,X_2]^C$, and hence also $\pi_*[Y_1,Y_2]=[\pi_*Y_1,\pi_*Y_2]$. 

{\it (iii)} For $Y=X^C\in\mathcal{S}$, we have $\xi^a\partial_a L+y^q\partial_q \xi^a\bar\partial_a L=0$. In the metric geometry limit $L(x,y)=g_{ab}(x)y^ay^b$, and hence $y^py^q(\xi^a\partial_ag_{pq}+g_{ap}\partial_q\xi^a+g_{aq}\partial_p\xi^a)=y^py^q\mathcal{L}_{X}g_{pq}(x)=0$. Since the Lie-derivative of the metric $g$ does not depend on the fibre coordinates of the tangent bundle we conclude $\mathcal{L}_{X}g_{pq}(x)=0$. This is the condition that defines $X$ as the symmetry generator of a metric spacetime. $\square$

\vspace{6pt}We now wish to study the implications of spherical, cosmological and maximal symmetry for the fundamental function $L$ of a Finsler spacetime.

\subsection{Spherical symmetry}\label{sec:sphsym}
Consider a Finsler spacetime $(M,L,F)$ and coordinates $(t,r,\theta,\phi,y^t,y^r,y^\theta,y^\phi)$ on its tangent bundle.
Spherical symmetry is defined by the following three vector fields, that generate spatial rotations and form the algebra $\mathfrak{so}(3)$,
\begin{eqnarray}
X_4=\sin\phi\partial_\theta+\cot\theta\cos\phi\partial_\phi\,,\quad 
X_5=-\cos\phi\partial_\theta+\cot\theta\sin\phi\partial_\phi\,,\quad
X_6=\partial_\phi\,.
\end{eqnarray}
Their complete lifts are obtained via the procedure described in the previous section
\begin{subequations}\label{eq:ssymc}
\begin{eqnarray}
X^C_4&=&\sin\phi\partial_\theta+\cot\theta\cos\phi\partial_\phi+y^\phi\cos\phi\bar\partial_\theta-\Big(y^\theta\frac{\cos\phi}{\sin^2\theta}+y^\phi\cot\theta\sin\phi\Big)\bar\partial_\phi\,,\\
X^C_5&=&-\cos\phi\partial_\theta+\cot\theta\sin\phi\partial_\phi+y^\phi\sin\phi\bar\partial_\theta-\Big(y^\theta\frac{\sin\phi}{\sin^2\theta}-y^\phi\cot\theta\cos\phi\Big)\bar\partial_\phi\,,\\
X^C_6&=&\partial_\phi \,.
\end{eqnarray}\end{subequations}
Applying the symmetry condition $X^C_6(L)=0$ implies $\partial_\phi L=0$, while using $X^C_4(L)=0$ and $X^C_5(L)=0$ to deduce  $(\sin\phi X^C_4-\cos\phi X^C_5)(L)=0$ and $(\cos\phi X^C_4+\sin\phi X^C_5)(L)=0$ yields
\begin{eqnarray}\label{eq:spherical}
\partial_\theta L=y^\phi \cot \theta\bar\partial_\phi L\,,\quad
y^\phi\sin^2\theta\bar\partial_\theta L=y^\theta\bar\partial_\phi L\,.
\end{eqnarray}
In order to analyze the implications of these equations on $L$ we introduce new coordinates
\begin{equation}\label{eq:c1}
 u(\theta)=\theta\,,\quad v(y^\theta)=y^\theta\,,\quad w(\theta, y^\theta,y^\phi)^2=(y^\theta)^2+\sin^2\theta (y^\phi)^2\,,
\end{equation}
while keeping $(t,y^t,r,y^r,\phi)$. The associated transformation of the derivatives
\begin{subequations}
\begin{eqnarray}
\partial_t&=&\partial_t\,,\quad \partial_r=\partial_r\,,\quad \partial_\theta=\frac{w^2-v^2}{w}\cot u\partial_w+\partial_u\,,\quad
\partial_\phi=\partial_\phi\,,\\
\bar\partial_t&=&\bar\partial_t\,,\quad \bar\partial_r=\bar\partial_r\,,\quad \bar\partial_\theta=\frac{v}{w}\partial_w+\partial_v\,,\quad\bar\partial_\phi=\sin u \frac{\sqrt{(w^2-v^2)}}{w}\partial_w\,,
\end{eqnarray}
\end{subequations}
makes the equations (\ref{eq:spherical}) equivalent to the simple constraints $\partial_u L=0$ and $\partial_v L=0$. 

Hence we conclude from the analysis of the symmetry conditions $X^C_i(L)=0$ that the most general spherically symmetric Finsler spacetime is described by a fundamental function which is $n$-homogeneous in $(y^t,y^r,w)$ and of the form
\begin{eqnarray}
L(t,r,\theta,\phi,y^t,y^r,y^\theta,y^\phi)=L(t, r,y^t,y^r,w(\theta,y^\theta,y^\phi))\,,
\end{eqnarray}
where $w(\theta,y^\theta,y^\phi)$ is defined in~(\ref{eq:c1}).

\subsection{Cosmological and maximal symmetry}\label{sec:cmsymm}
After our discussion of the spherically symmetric case in full detail above, we will now present the results of a similar analysis first for cosmologically and second for maximally symmetric Finsler spacetimes.

Cosmological symmetry describes an isotropic and homogeneous spacetime. This is a much more symmetric situation than in the spherically symmetric scenario, and is implemented by requiring the following six vector fields to be symmetry generators, see~\cite{Garecki},
\begin{subequations}\label{eq:csym}
\begin{eqnarray}
X_1&=&\chi\sin\theta \cos\phi
\partial_r+\frac{\chi}{r}\cos\theta\cos\phi\partial_\theta-\frac{\chi}{r}\frac{\sin\phi}{\sin\theta}\partial_\phi\,,\\
X_2&=&\chi\sin\theta \sin\phi
\partial_r+\frac{\chi}{r}\cos\theta\sin\phi\partial_\theta+\frac{\chi}{r}\frac{\cos\phi}{\sin\theta}\partial_\phi\,,\\
X_3&=&\chi\cos\theta \partial_r-\frac{\chi}{r}\sin\theta\partial_\theta\,,\\
X_4&=&\sin\phi\partial_\theta+\cot\theta\cos\phi\partial_\phi\,,\quad
X_5=-\cos\phi\partial_\theta+\cot\theta\sin\phi\partial_\phi\,,\quad
X_6=\partial_\phi\,,
\end{eqnarray}
\end{subequations}
where we write $\chi=\sqrt{1-kr^2}$ and $k$ is constant. The complete lifts of these vector fields are listed in appendix~\ref{app:compl}. Applying the symmetry conditions $X_i^C(L)=0$ to the fundamental function $L$ and introducing the new coordinates
\begin{subequations}
\begin{eqnarray}
&&q(r)=r\,,\quad s(y^r)=y^r\,,\quad u(\theta)=\theta\,,\quad v(y^\theta)=y^\theta\,,\\
&& w_C(r,\theta,y^r,y^\theta,y^\phi)^2=\frac{(y^r)^2}{1-kr^2}+r^2\big((y^\theta)^2+\sin^2\theta (y^\phi)^2\big) ,
\end{eqnarray}
\end{subequations}
while keeping $(t,y^t)$ yields the following result: the cosmological fundamental function $L$ is $n$-homogeneous in $(y^t,w_C)$ and has the form
\begin{equation}
L(t,r,\theta,\phi,y^t,y^r,y^\theta,y^\phi)=L(t,y^t,w_C(r,\theta,y^r,y^\theta,y^\phi))\,.
\end{equation}
The constant $k$ only appears in the expression for the coordinate $w_C$. The value of $w_C$ can be understood as the metric length measure on a three-dimensional manifold of constant curvature~$k$. The same metric appears in the spatial part of the standard Robertson--Walker metric.

For the study of maximally symmetric Finsler spacetimes we use some notation from~\cite{Weinberg}, where such spacetimes are constructed from embeddings into a five-dimensional manifold. Symmetry vectors generating maximal symmetry are given by 
\begin{equation}
 X_\alpha=C(x)\alpha^c\partial_c\,,\quad X_\Omega=\Omega^a{}_bx^b\partial_a\,,
\end{equation}
with $C(x)=\sqrt{1-K C_{pq}x^px^q}$, constant $K$, and constant $4\times4$ matrices $C_{ab}$ and $\Omega^a{}_b$. There are four linearly independent vector fields $X_\alpha$ and six $X_{\Omega}$ by requiring the condition $\Omega^q{}_bC_{qa}=-\Omega^q{}_aC_{qb}$; their complete lifts are
\begin{equation}
 X^C_{\alpha}=C(x)\alpha^c\partial_c-y^b\frac{KC_{bm}x^m}{C(x)}\alpha^c\bar\partial_c\,,
 \quad X^C_{\Omega}=\Omega^a{}_bx^b\partial_a+y^b\Omega^a{}_b\bar\partial_a\,.
\end{equation}
Evaluating the symmetry conditions $X_\alpha^C(L)=0$ and $X_\Omega^C(L)=0$ on the fundamental function, and introducing new coordinates 
\begin{equation}\label{eq:c3}
u^a(x)=x^a\,,\quad v^\gamma(y)=y^\gamma\,,\quad 
w_M(x,y)^2=C_{ab}y^ay^b+\frac{K}{C(x)^2}C_{ap}x^py^aC_{bq}x^qy^b=g_{ab}(x)y^ay^b\,,
\end{equation}
where $\gamma$ runs over any three indices in $\{0,1,2,3\}$, yields the following result: the maximally symmetric  fundamental function $L$ is $n$-homogeneous in $w_M$, and of the form
\begin{equation}
L(x,y)=L(w_M(x,y))=A\ w_M(x,y)^n\,.
\end{equation}
The final equality is obtained from Euler's theorem for homogeneous functions. 

Observe that the maximally symmetric fundamental function always describes a metric geometry, see~(\ref{eq:c3}). Hence all maximally symmetric Finsler spacetimes are Lorentzian spacetimes, and the gravity equation~(\ref{eq:complGrav}) is equivalent to Einstein's equations. Thus we can immediately conclude that the only maximally symmetric, source free vacuum solution of our Finsler gravity equation is the Minkowski metric induced fundamental function $L=\eta_{ab}y^ay^b$ and its powers. In the expression for $w_M$ above this corresponds to $C_{ab}=\eta_{ab}$ and $K=0$. Maximally symmetric spacetimes with $K\neq 0$ can only be obtained as solutions of the Finsler gravity equation by adding a cosmological constant term, similarly as in general relativity.

\section{Lowest order effects in the solar system}\label{sec:Corr}
In this section we will study Finsler spacetimes that describe mild deviations from Lorentzian geometry. In this situation, the complicated Finsler gravity field equation allows a simplified treatment. After a general discussion of the linearised  field equation, we will employ what we learned about spacetime symmetries to present a spherically symmetric solution. This particular model turns out to be a refinement of the linearised Schwarzschild solution of general relativity, and we will argue that it should be capable of modelling unexplained effects in the solar system like the fly-by anomaly.

\subsection{Finsler modifications of Lorentzian geometry}\label{subsec:ModLGeom}
Recall that the fundamental functions $L=L_0$ and $L=(L_0)^k$ define the same geometry, and that this is respected by the Finsler gravity field equation. Hence the following class of fundamental functions gives us good control over deviations from Lorentzian metric geometry, 
\begin{equation}\label{eq:Lpert}
L=\big(g_{ab}(x)y^ay^b\big)^k+h(x,y)=G(x,y)^k+h(x,y)\,.
\end{equation} 
Here, $h(x,y)$ is a $2k$-homogeneous function that causes the Finsler modifications of the Lorentzian metric spacetime $(M,g)$. The abbreviation $G(x,y)$ should not be confused with the Sasaki-type metric on $TM$. 

Recall the Finsler gravity vacuum field equation from~(\ref{eq:gravvac}):
\begin{equation}\label{eq:gravvac2}
 g^{F\,ab}\bar\partial_a\bar\partial_b \mathcal{R}-\frac{6}{F^2}\mathcal{R}+2 g^{F\,ab}(\nabla_aS_b+S_aS_b+\bar\partial_a\nabla S_b)=0\,.
\end{equation}
We will now expand this equation to linear order in the modification $h(x,y)$, where $G(x,y)\neq 0$. In the following calculations we suppress all higher order terms. We introduce $l=\frac{G^{1-k}}{k}h$ and $l_{ab}=\frac{1}{2}\bar\partial_a\bar\partial_b l$ to expand the Finsler function and Finsler metric as
\begin{equation}\label{eq:F2}
F^2\simeq \frac{G}{|G|}(G+l)\,,\quad g^F_{ab}\simeq \frac{G}{|G|}(g_{ab}+l_{ab})\,.
\end{equation}
The coefficients of the nonlinear connection are calculated from equation (\ref{eq:nonlin}):
\begin{equation}\label{eq:nlinfirst}
N^a{}_b\simeq y^m\Gamma^{a}{}_{bm}-\frac{1}{2}g_{cq}\Gamma^c{}_{mn}y^my^n\bar\partial_bl^{aq}- l^{aq}g_{cq}\Gamma^c_{bm}y^m+\frac{1}{2}g^{aq}(\partial_bl_{qm}+\partial_ml_{qb}-\partial_ql_{bm})y^m\,.
\end{equation}
Here, the $\Gamma^{a}{}_{bc}$ are the Christoffel symbols of the metric $g$, and in the following $\nabla$ acts as the Levi--Civita connection. Note that the zeroth order term, for $l\rightarrow 0$, is the metric linear connection. The curvature and the tensor $S$ can be expressed with help of the shorthand notation
\begin{equation}
T^{ a}{}_{bc}=\frac{1}{2}g^{aq}\big(\nabla_bl_{qc}+\nabla_cl_{qb}-\nabla_ql_{bc}\big)
\end{equation}
as
\begin{subequations}
\begin{eqnarray}
\mathcal{R}&=&R^a{}_{ab}y^b \simeq -y^ay^bR_{ab}[g]- \nabla_a (y^by^cT^a{}_{bc})+ \nabla(y^cT^a{}_{ac})\,,\\
S_a&=&\Gamma^{\delta p}{}_{pa}-\bar\partial_p N^p{}_a \simeq -y^q\bar\partial_aT^p{}_{pq}\,.
\end{eqnarray}
\end{subequations}
The zeroth order term in $\mathcal{R}$ is determined by the Ricci tensor of $g$, while $S_a\rightarrow 0$. Collecting all terms in the gravitational field equation (\ref{eq:gravvac2}) finally yields
\begin{eqnarray}\label{eq:line}
0 \simeq &-&2\frac{G}{|G|}R[g]+\frac{6}{G}y^ay^bR_{ab}[g]\\
&+&\Big[2 \frac{G}{|G|} l^{ab}R_{ab}[g]+\frac{6G-2l}{G^2}y^ay^bR_{ab}[g]+g^{ab}\bar\partial_a\bar\partial_b(-\nabla_a (y^by^cT^a{}_{bc})+\nabla(y^cT^a{}_{ac}))\nonumber \\
&&\quad- \frac{6}{G} (-\nabla_a (y^by^cT^a{}_{bc})+\nabla(y^cT^a{}_{ac}))-2 g^{ab}(\nabla_ay^q\bar\partial_bT^p{}_{pq}+\bar\partial_a\nabla y^q\bar\partial_bT^p{}_{pq})\Big]\,.\nonumber
\end{eqnarray}
The zeroth order contribution in the first line is equivalent to the Einstein vacuum equations, as discussed in section~\ref{sec:gravvac}. The first order terms in square brackets determine the Finsler modification of the unperturbed metric  background solution. The details of how to rewrite the different terms of this equation in terms of the perturbation $h$ in the fundamental function $L$, see~(\ref{eq:Lpert}), instead of $l$ can be found in appendix~\ref{app:lin}.

\subsection{Refinements to the linearised Schwarzschild solution}\label{subsec:linSchw}
We will now use our results on symmetries and on the linearisation of vacuum Finsler gravity around metric spacetimes to derive a particular model that refines the linearised Schwarzschild solution and can be used to study solar system physics.

Recall from section~\ref{sec:sphsym} that the dependence of the general spherically symmetric fundamental function in tangent bundle coordinates induced by $(t,r,\theta,\phi)$ is restricted to $L(t,r,y^t,y^r,w(\theta,y^\theta,y^\phi))$ where $w^2=(y^\theta)^2+\sin^2\theta (y^\phi)^2$. We wish to study such a spherically symmetric fundamental function that describes a Finsler modification of Lorentzian geometry. For simplicity, we consider a bimetric four-homogeneous Finsler spacetime that perturbs the maximally symmetric vacuum solution of Finsler gravity which is given by Minkowski spacetime. We assume $L=\big(\eta_{ab}y^ay^b\big)^2+\eta_{ab}y^ay^bh_{cd}y^cy^d = (\eta_{ab}y^ay^b)(\eta_{cd}+h_{cd})y^cy^d$ with $h_{ab}=\mathrm{diag}(a(r),b(r),c(r)r^2,c(r)r^2\sin^2\theta)$. This ansatz has the explicit form
\begin{equation}\label{eq:bisphL}
L(r,y^t,y^r,w)=\big(-y^{t^2}+y^{r^2}+r^2w^2\big)\big([-1+a(r)]y^{t^2}+[1+b(r)]y^{r^2}+[1+c(r)]r^2w^2\big).
\end{equation}

Observe that the function $c(r)$ cannot be transformed away by defining a new radial coordinate. Although this could remove $c(r)$ from the metric in the right hand bracket, such a coordinate change would generate extra terms in the metric appearing in the left hand bracket. Therefore, the existence of the function $c(r)$ as a physical degree of freedom is a Finsler geometric effect that appears as a consequence of the bimetric spacetime structure assumed here.

We will now solve the linearised Finsler gravity equation~(\ref{eq:line}) for $a(r)$, $b(r)$ and $c(r)$ with the ansatz~(\ref{eq:bisphL}). Sorting the equation with respect to powers in $y^t,\ y^r$ and $w$ gives rise to three equations that have to be satisfied:
\begin{eqnarray}
-2 a'-ra''=0\,,\quad ra''+2b'-4c'-2rc''=0\,,\quad  ra'+2b+rb'-2c-4rc'-r^2c''=0\,.
\end{eqnarray}
The solution of these equations is
\begin{equation}
 a(r)=-\frac{A_1}{r}+A_2,\quad b(r)=-\frac{A_1}{r}+\frac{A_3}{r^2},\quad c(r)=\frac{A_4}{r}-\frac{A_3}{r^2}\,.
\end{equation}

We will now study the properties of this specific first order Finsler spacetime solution and compare it to the linearized Schwarzschild spacetime. We use the linearised expression for the non-linear connection coefficients in~(\ref{eq:nlinfirst}) to analyze the Finsler geodesic equation that is derived by extremizing the proper time integral~(\ref{eq:Fact}). For a curve with coordinates $x(\tau)$ this has the form $\ddot x^a + N^a{}_b(x,\dot x)\dot x^b=0$. As usual in spherical symmetry, setting $\theta=\frac{\pi}{2}$ solves one of the four component equations immediately; the remaining equations are
\begin{subequations}
\begin{eqnarray}
0&=&\ddot t-\frac{1}{2}\frac{A_1}{r^2}\ \dot t\ \dot r\\
0&=&\ddot r -\frac{1}{4}\frac{A_1}{r^2}\ \dot t^2+\frac{1}{4}\Big(\frac{A_1}{r^2}-2\frac{A_3}{r^3}\Big)\ \dot r^2+\Big(-r-\frac{A_1}{2}-\frac{A_4}{4}+\frac{1}{2}\frac{A_3}{r}\Big)\ \dot \phi^2\\
0&=&\ddot \phi+\frac{2}{r}\Big( 1-\frac{1}{4}\frac{A_4}{r}+\frac{1}{2}\frac{A_3}{r^2}\Big)\ \dot\phi \ \dot r\,.
\end{eqnarray}
\end{subequations}
From these equations we find two constants of motion
\begin{equation}
 E=\dot t\ \Big(1+\frac{1}{2}\frac{A_1}{r}\Big), \quad 
 \ell=r^2 \Big(1+\frac{1}{2}\frac{A_4}{r}-\frac{1}{2}\frac{A_3}{r^2}\Big)\dot\phi\,.
\end{equation}
These can be used to deduce the orbit equation from the affine normalization condition that $F(x,\dot x)=1$ along the Finsler geodesic; we employ~(\ref{eq:F2}) and write $\sigma$ for the sign of $\eta_{ab}\dot x^a\dot x^b=- \dot t^2+\dot r^2+r^2 \dot\phi^2$ to obtain
\begin{equation}
\frac{1}{2}\dot r ^2= \frac{E^2}{2}\Big(1-\frac{A_2}{2}\Big)+\frac{1}{2}\sigma\Big(1+\frac{A_1}{2r}\Big)-\frac{\ell^2}{2r^2}\Big(1+\frac{A_1}{2r}-\frac{A_4}{2r}\Big)+\frac{A_3}{4 r^2}\Big(\sigma-E^2\Big).
\end{equation}

The geodesic equations, the constants of motion and the orbit equation are well suited to compare the bimetric linearised Finsler solution with the linearised Schwarzschild solution. To see the differences to this solution of Einstein gravity we first note that $A_2$ can be absorbed into a redefinition of $E$, hence can be assumed to be zero. Second we introduce the Schwarzschild radius $r_0$ to redefine $A_1=-2r_0 (1+a_1)$, $A_3=2\ell^2 a_3/(E^2-\sigma)$ and $A_4=2r_0a_4$ in terms of dimensionless small constants $a_1$, $a_3$ and $a_4$. Then the orbit equation becomes
\begin{equation}
\frac{1}{2}\dot r ^2= \frac{E^2}{2} +\frac{\sigma}{2} -\frac{\sigma r_0}{2r} (1+a_1) - \frac{\ell^2}{2r^2}(1+a_3) + \frac{r_0 \ell^2}{2r^3} (1+a_1+a_4)\,.
\end{equation}
In the special case $a_1=a_3=a_4=0$ this is precisely the orbit equation in the linearized Schwarzschild geometry, see~\cite{Carroll}; the same limit also applies to the geodesic equations and the constants of motion. 

\begin{figure}[h]
\includegraphics[width=0.5\textwidth]{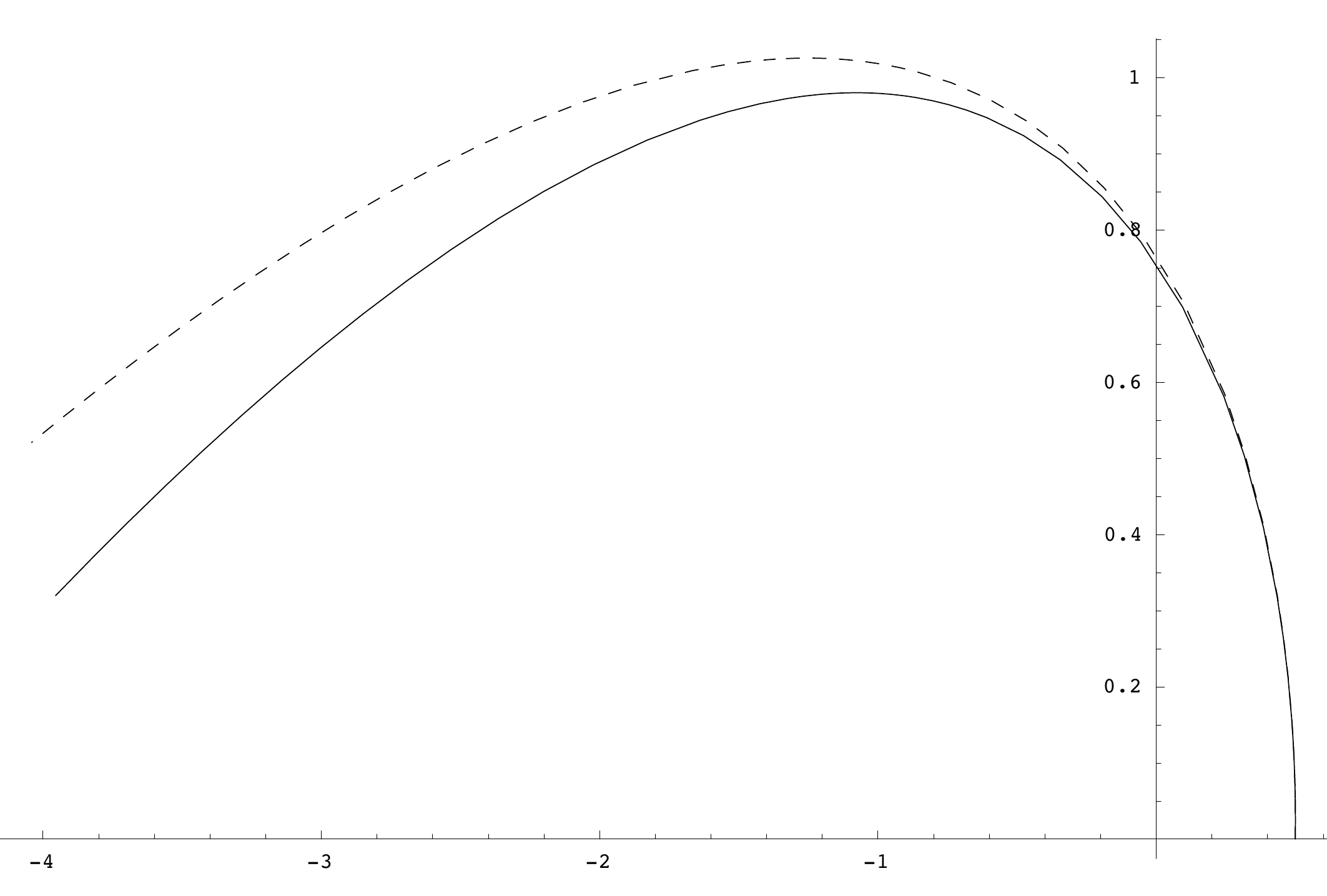}
\caption{\label{fig:flyby}\textit{Numerical fly-by solutions of the geodesic equations for linearized Schwarzschild geometry (dashed line) and the bimetric Finsler refinement (solid line) with $a_1=0$, $a_3\simeq 0.156$ and $a_4=0.1$. The mass is centred at the origin an has Schwarzschild radius $r_0 = 0.1$. The initial conditions are $r(0)=0.5$, $\dot r(0)=0.02$, $\phi(0)=0$, $\dot\phi(0)=1.1$ and $t(0)=0$ for both curves, and $\dot t(0)$ is calculated from the respective unit normalization condition $F(x,\dot x)=1$.}}
\end{figure}

The Finsler geometric refinements of the metric Schwarzschild geometry are encoded in the constants $a_1, a_3$ and $a_4$. These can in principle be fitted to data from solar system experiments. Indeed, there are certain observations that cannot be fully explained by the Schwarzschild solution~\cite{Lammerzahl:2006ex}, for instance, the fly-by anomaly: for several spacecrafts it has been reported that swing-by manoeuvres lead to a small unexplained velocity increase. This corresponds to a change in the shape of the orbit of the spacecraft. Such a change can in principle be modelled by Finsler refinements; the perturbations $a_1$, $a_3$ and $a_4$ certainly provide possibilities to alter the wideness of the swing-by orbit as compared to that expected from Einstein gravity. This can be confirmed by simple numerical calculations, see figure~\ref{fig:flyby}.

We have seen that Finsler geometries exist that are extremely close to metric geometries. Our specific example of a spherically symmetric bimetric perturbation around Minkowski spacetime could be reinterpreted as a geometry close to the linearized Schwarzschild solution of Einstein gravity. The more complex causal structure, however, leads to additional constants that modify the geodesic equations and in particular the shape of test particle orbits. This could be a means to explain the fly-by anomaly in the solar system. We emphasize that this consequence already at first order perturbation theory gives a glimpse on the potential of Finsler gravity.

\section{Discussion}\label{sec:disc}
Finsler geometry is fundamentally based on the reparametrization invariant length integral~(\ref{eq:Fact}). In physics, this integral can be used as a very general  clock postulate on the one hand, and on the other as an action for massive point particles which automatically guarantees the very precisely tested weak equivalence principle. In previous work~\cite{Pfeifer:2011tk} we formulated a set of minimal requirements for the application of Finsler geometry to the description of spacetime. This led us to Definition~1 of Finsler spacetimes $(M,L,F)$ which have sufficient structure to provide notions of causality and are mathematically controlled generalizations of Lorentzian geometry.

In this article, we constructed an action for Finsler gravity from first principles. Our theory of Finsler gravity~(\ref{eq:compact}) fully includes the description of matter fields which are coupled to Finsler spacetime by a lifting principle that generates the appropriate action from the standard Lagrangian on Lorentzian spacetime. We derived the gravitational field equation by variation with respect to the fundamental geometry function~$L$, and could show that it consistently becomes equivalent to the Einstein field equations in the metric geometry limit. Hence Einstein gravity can be seen as special case of our gravity theory based on Finsler geometry. We presented the geometric Definition~2 of observers on Finsler spacetimes along with a clear interpretation of how they measure physical fields. By Theorem~2 we were able to characterize the class of transformations that relates two different observers at the same point of the spacetime manifold; these transformations have the algebraic structure of a groupoid as proven in Theorem~3. Any observer transformation can be understood as the composition of a usual Lorentz transformation and an identification of the two observers' four-velocities by a geometrically well-defined parallel transport. For the limiting case of metric geometries, this parallel transport trivializes so that the transformation groupoid becomes equivalent to the standard Lorentz group. In this sense, Finsler spacetimes generalize Lorentz invariance instead of violating it.

As a further formal development we presented Definition~4 of symmetries of Finsler spacetimes and Theorem~4 that shows some of their basic properties. We applied this notion to determine the most general fundamental geometry functions consistent with spherical, cosmological, or maximal symmetry. Maximally symmetric Finsler spacetimes are in fact maximally symmetric metric spacetimes, and the only maximally symmetric source-free vacuum solution of Finsler gravity is Minkowski spacetime. As a concrete application of the results of this article we studied a simple spherically symmetric bimetric perturbation around this flat metric vacuum. We found a first order solution of the Finsler gravity equation which is a refinement of the linearized Schwarzschild solution of Einstein gravity. With a special choice of the parameters in our solution the resulting geodesics are identical to those of linearized Schwarzschild spacetime. This demonstrates that weak field gravitational experiments may not be sufficient to distinguish Finsler spacetimes from Lorentzian metric spacetimes. But we saw that the full set of parameters in our model solution could be capable to resolve the fly-by anomaly in the solar system.
 
This is very promising, and creates a strong motivation for more intensive studies of our new theory of gravity. It is natural to ask whether the additional degrees of freedom of Finsler spacetime solutions as compared to metric spacetimes may lead to new insights on the dark matter distributions in galaxies or dark energy in the universe. These could be effects of a fundamentally more complex spacetime geometry instead of being particle physics phenomena. It will be possible to study these questions once solutions of Finsler gravity, especially for spherical symmetry and cosmology, become available. For cosmology, it will also be necessary to study perfect fluid sources for Finsler spacetimes.

We saw in~\cite{Pfeifer:2011ve} that Finsler spacetimes can provide a geometric explanation of the OPERA measurements of superluminal neutrinos~\cite{Adam:2011zb}. The velocity difference between the neutrinos and the speed of light recognized as the boundary velocity of observers depends not only on the energy and mass of the neutrino, but also on the underlying spacetime geometry. So also in this context, solutions for spherical symmetry and cosmology are needed in order to understand the size of the effect for the different observed neutrino sources.

Further important topics for future research are the coupling of spinor fields, the analysis of field theories on the generalized causal structure of Finsler spacetimes and their quantization. 

\acknowledgments CP and MNRW thank Niklas H\"ubel, Claudio Dappiaggi, Andreas Degner, Klaus Fredenhagen, Manuel Hohmann, Matthias Lange, Falk Lindner, Gunnar Preuss and Felix Tennie for inspiring discussions. They acknowledge full financial support from the German Research Foundation DFG under grant WO 1447/1-1.

\appendix
\section{Technical details}\label{app:details}
This appendix presents technical details for several derivations in the main text. In particular, we show how to perform the variation of the Finsler gravity and matter field actions on the unit tangent bundle; for completeness we state the complete lifts of the cosmological symmetry generators; and we display some additional material on the linearized Finsler gravity equations.

\subsection{Variation of the gravity action}\label{app:vargrav}
The Finsler gravitational field equation presented in section~\ref{sec:gravvac} can be deduced from our new Finsler gravity action as follows. Before we consider the variation of the matter part with respect to the fundamental geometry function $L$ in the next section, we here take a look at the pure gravity action~(\ref{eq:gravact}):
\begin{equation}
 S_G[L]=\int d^4\hat xd^3u \left[\sqrt{g^Fh^F}\ R^a{}_{ab}y^b\right]_{|\Sigma}\,.
\end{equation}

The integrand is homogeneous of degree five; to obtain the first intermediate step~(\ref{eq:calcgrav1}) of the variation we use the facts that for $f(x,y)$ homogeneous of degree~$k$ holds $f(x,y)_{|\Sigma}=\frac{f(x,y)}{F(x,y)^k}$ and that $\delta_L(f(x,y)_{|\Sigma})=(\delta_Lf(x,y))_{|\Sigma}-\frac{k}{n}f(x,y)_{|\Sigma}\frac{\delta L}{L}$.

The second step~(\ref{eq:calcgrav2}) is obtained by using the coordinate transformation formulae~(\ref{eq:invvv}) and the fact that $\delta(\partial_\alpha y^a)=-y^a\partial_\alpha (\frac{\delta L}{n L})$ to calculate 
\begin{eqnarray}
h^{F\alpha\beta}\delta h^F{}_{\alpha\beta}=(g^{F ab}\bar\partial_a u^\alpha\bar\partial_b u^\beta)(\partial_\alpha y^c\partial_\beta y^d \delta g^F{}_{cd}+2\partial_\alpha \delta y^c\partial_\beta y^d g^F_{cd})=g^{F ab}\delta g^F_{ab}-\frac{2}{n}\frac{\delta L}{L}\,,
\end{eqnarray}
which in turn is used to deduce
\begin{equation}\label{eq:volu}
 \delta (\sqrt{g^Fh^F}\ R^a{}_{ab}y^b)=\sqrt{g^Fh^F}\Big(\Big[g^{F ab}\delta g^F{}_{ab}-\frac{1}{n}\frac{\delta L}{L}\Big]R^a{}_{ab}y^b+\delta R^a{}_{ab}\,y^b\Big).
\end{equation}

The formulae (\ref{eq:calcgrav}) used in the third step of the variation are basically obtained by means of integration by parts~(\ref{eq:intbp}). For a function $f(x,y)$ that is $k$-homogeneous in $y$ the following holds
\begin{equation}
\int d^4\hat xd^3u \left(\sqrt{g^Fh^F} g^{F ab}\delta g^F_{ab} f \right)_{|\Sigma}=
\int d^4\hat xd^3u \Big(\sqrt{g^Fh^F} \big(f(k+4)(2-k)+F^2 g^{F ab}\bar\partial_a\bar\partial_b f\big)\frac{\delta L}{nL} \Big)_{|\Sigma}\,;
\end{equation}
choosing $f=R^a{}_{ab}y^b$ which has $k=2$ proves formula~(\ref{eq:calcgrav31}). To show equation~(\ref{eq:calcgrav32}) we first write $S^a{}_{bc}=-y^q\bar\partial_b\Gamma^{\delta a}{}_{qc}$ and use
\begin{equation}
 \delta R^a{}_{bc}=-2y^d\nabla_{[b}\delta \Gamma^{\delta a}{}_{c]d}+2y^pS^a{}_{q[b}\Gamma^{\delta q}{}_{c]p}
\end{equation}
to equate
\begin{eqnarray}
\delta R^a{}_{ab}\,y^b&=&-2y^by^q\Big(\nabla_{[a}\delta \Gamma^{\delta a}{}_{b]q}-\frac{1}{2}S_{c}\delta\Gamma^{\delta c}{}_{bq}\Big)\nonumber\\
&=&-\nabla_a (y^by^q\delta \Gamma^{\delta a}{}_{bq})+y^b\nabla_b(y^q\delta \Gamma^{\delta a}{}_{aq})+S_{c}\delta\Gamma^{\delta c}{}_{bq}y^by^q\,.
\end{eqnarray}
The integration by parts formulae~(\ref{eq:intbp}) and
\begin{eqnarray} 
y^by^q\delta\Gamma^{\delta a}{}_{bq}&=&\frac{1}{2}g^{L ap}(y^b\nabla_b\bar\partial_p\delta L-\nabla_p \delta L)\nonumber\\
&=&\frac{|L|^{2/n}}{n L}g^{Fab}(y^b\nabla_b\bar\partial_p\delta L-\nabla_p \delta L)+\frac{(2-n)}{n L}y^ay^b\nabla_b\delta L
\end{eqnarray}
then yield the desired equation
\begin{eqnarray}
 \int d^4\hat xd^3u \left(\sqrt{g^Fh^F} \delta R^a{}_{ab}y^b \right)_{|\Sigma}=\int d^4\hat xd^3u \left(\sqrt{g^Fh^F}\ 2\ S_{c}\delta\Gamma^{\delta c}{}_{bq}y^by^q \right)_{|\Sigma}\nonumber\\
 =\int d^4\hat xd^3u \Big(\sqrt{g^Fh^F}\ 2\ F^2g^F_{ab}\big(\nabla_aS_b+S_aS_b+\bar\partial_a\nabla S_b\big)\frac{\delta L}{nL}\Big)_{|\Sigma}\,.
\end{eqnarray}
Combining these three steps as we did in section~\ref{sec:gravvac} finally produces the Finsler gravity vacuum field equation~(\ref{eq:gravvac}).

\subsection{Variation of the matter action}\label{app:varmat}
In section~\ref{sec:gravmat} we presented a coupling principle of matter fields to Finsler gravity. The crucial steps of the derivation of the constraints~(\ref{eq:cons}), equations of motion~(\ref{eq:eom1}) and~(\ref{eq:eom2}), and of the metric limit of the complete gravity equation~(\ref{eq:fgrav}) including the matter source terms shall be presented here. Recall the matter action for a $p$-form field $\Phi(x,y)$ on Finsler spacetime arises from a lift of the standard $p$-form action on Lorentzian spacetime as
\begin{equation}\label{eq:actm}
S_m[L,\Phi, \lambda]=\int_\Sigma d^4xd^3u\ \Big[\sqrt{g^Fh^F} \Big(\mathcal{L}(G,\Phi, \mathrm{d}\Phi)+\lambda (1-P^H)\Phi\Big) \Big]_{|\Sigma}\,.
\end{equation}
In order to perform the variation we consider all objects in the horizontal/vertical basis of $TTM$ where $G$ is diagonal, see~(\ref{eq:sasaki}). In the following the $M,N,..$ label both horizontal and vertical indices, $\bar a,\bar b,...$ label vertical indices, and $a,b...$ label horizontal indices. Then
\begin{equation}
\mathcal L(G, \Phi, \mathrm{d}\Phi)+\lambda (1-P^H)\Phi=\mathcal{L}(G_{MN}, \Phi_{M_1...M_p}, \mathrm{d}\Phi_{AM_1...M_p})+\lambda^{\bar a_1M_2...M_p}\Phi_{\bar a_1 M_2..M_p}\,,
\end{equation}
and the variation of this Lagrangian can now be written as follows
\begin{eqnarray}\label{eq:vari}
\delta (\mathcal{L}+\lambda (1-P^H)\Phi)&=&\frac{\partial \mathcal{L}}{\partial G_{MN}}\delta G_{MN}+\frac{\partial \mathcal{L}}{\partial \Phi_{M_1...M_p}}\delta \Phi_{M_1M_p}+\frac{\partial \mathcal{L}}{\partial (\mathrm{d}\Phi_{NM_1...M_p})}\delta (\mathrm{d}\Phi_{NM_1M_p})\nonumber\\
&&{}+\lambda^{\bar a_1M_2...M_p}\delta\Phi_{\bar a_1 M_2..M_p}+\delta\lambda^{\bar a_1M_2...M_p}\Phi_{\bar a_1 M_2..M_p}\,.
\end{eqnarray}

We can immediately read off the variation with respect to the Lagrange multiplier components which produces~(\ref{eq:cons}). Hence the Lagrange multiplier $\lambda$ sets to zero all components of $\Phi$ with at least one vertical index, so that only purely horizontal components remain on-shell. 

The expansion of $\mathrm{d}\Phi$ in components with respect to the horizontal/vertical basis yields
\begin{equation}
\mathrm{d}\Phi_{NM_1...M_p}=(p+1)D_{[N}\Phi_{M_1...M_p]} - \frac{p(p+1)}{2}\gamma^Q{}_{[NM_1}\Phi_{|Q|M_2...M_p]}\,,
\end{equation}
where we write $D_M=\delta^a_M \delta_a + \delta^{\bar a}_M \bar\partial_a$, and $\gamma^Q{}_{MN}$ denote the commutator coefficients of the horizontal/vertical basis. Their only non-vanishing components are given by $\gamma^{\bar a}{}_{bc}=[\delta_b,\delta_c]^{\bar a}=R^{\bar a}{}_{bc}$ and $\gamma^{\bar a}{}_{\bar b c}=[\bar\partial_b,\delta_c]^{\bar a}=\bar\partial_b N^{\bar a}{}_c$. One now uses the integration by parts formulae~(\ref{eq:intbp}) to obtain the variation of the matter action with respect to $\Phi$; this produces the equations of motion~(\ref{eq:eom1}) and~(\ref{eq:eom2}).

Finally the source term for the gravity field equation is obtained by variation of the matter action $S_m$ in~(\ref{eq:actm}) with respect to the fundamental geometry function $L$. This not only includes the variation~(\ref{eq:vari}) but also that of the volume element which can be read off from~(\ref{eq:volu}). We will now show that the metric limit of Finsler gravity plus matter is consistent; this can be done on-shell where we may use the Lagrange multiplier constraints to set all explicitly appearing $\Phi_{\bar a_1 M_2..M_k}$ to zero. Then the variation of $S_m$ with respect to $L$ becomes
\begin{eqnarray}
\delta S_m[L,\Phi]&=&\int_\Sigma d^4xd^3u\ \sqrt{g^Fh^F}_{|\Sigma} \Big[\big(g^{Fab}\bar\partial_a\bar\partial_b\mathcal{L}+4\mathcal{L}\big)\frac{\delta L}{nL}\nonumber\\
&&\qquad\qquad\qquad\qquad\qquad+\frac{\partial \mathcal{L}}{\partial G_{MN}}\delta G_{MN}+\frac{\partial \mathcal{L}}{\partial (\mathrm{d}\Phi_{NM_1...M_k})}\delta (\mathrm{d}\Phi_{NM_1M_k})\Big]_{|\Sigma}\nonumber\\
&=&\int_\Sigma d^4xd^3u\ \sqrt{g^Fh^F}_{|\Sigma} \Big[\big(g^{Fab}\bar\partial_a\bar\partial_b\mathcal{L}+4\mathcal{L}\big)\frac{\delta L}{nL}\\
&&\qquad\qquad\qquad\qquad\qquad+\frac{\partial \mathcal{L}}{\partial g^F_{ab}}\delta g^F_{ab}+\frac{\partial \mathcal{L}}{\partial g^F_{\bar a\bar b}}\delta\Big( \frac{g^F_{\bar a \bar b}}{F^2}\Big)+\frac{\partial \mathcal{L}}{\partial (\mathrm{d}\Phi_{ba_1...a_k})}\delta (\delta_{[b}\Phi_{a_1..a_k]})\Big]_{|\Sigma}\,.\nonumber
\end{eqnarray}

In order to determine the energy momentum scalar $T_{|\Sigma}$ defined in~(\ref{eq:EMS}) on a generic Finsler spacetime one has to calculate all terms in the expression above carefully. However, in the metric geometry limit the last two terms vanish. Indeed, $\frac{\partial \mathcal{L}}{\partial g^F_{\bar a\bar b}}$ is always composed from terms with vertical indices that must be either of the type $\bar\partial\Phi$ or contain components of $\Phi$ with at least one vertical index; the last term is proportional to $\delta N \bar \partial \Phi$; in the metric limit $\bar\partial \Phi$ vanishes and the vertical index components of $\Phi$ are zero on-shell. Therefore the remaining terms that are relevant in the metric limit are
\begin{equation}
\delta S_m[L,\Phi] \rightarrow \int_\Sigma d^4xd^3u\ \sqrt{g^Fh^F}_{|\Sigma} \Big[\big(g^{Fab}\bar\partial_a\bar\partial_b\mathcal{L}+4\mathcal{L}\big)\frac{\delta L}{nL}+\frac{\partial \mathcal{L}}{\partial g^F_{ab}}\delta g^F_{ab}\Big]_{|\Sigma}\,.
\end{equation}
The rewriting $\delta g^F_{ab}=\frac{1}{2}\bar\partial_a\bar\partial_b \delta F^2$ and subsequent integration by parts yields
\begin{eqnarray}
&&\int_\Sigma d^4xd^3u\ \sqrt{g^Fh^F}_{|\Sigma} \Big[\frac{\partial \mathcal{L}}{\partial g^F_{ab}}\delta g^F_{ab}\Big]_{|\Sigma}\\
&=&\int_\Sigma d^4xd^3u\ \sqrt{g^Fh^F}_{|\Sigma} \Big[-\bar\partial_c K^c+\Big(-g^{Fij}\bar\partial_cg^F_{ij}+4g^F_{ic}y^i\Big)K^c\Big]_{|\Sigma}\frac{\delta L}{nL}\nonumber\,,
\end{eqnarray}
with 
\begin{equation}
K^c=\Big(-g^{Fij}\bar\partial_d g^F_{ij}+\frac{4}{F^2}g^F_{id}y^i\Big)\frac{\partial \mathcal{L}}{\partial g^F_{cd}}-\bar\partial_d\frac{\partial \mathcal{L}}{\partial g^F_{cd}}\,.
\end{equation}
Applying the metric limit now means to consider $L(x,y)=g_{ab}(x)y^ay^b$ with the consequence that $g^F_{ab}(x,y)=-g_{ab}(x)$ for timelike $y$. The expression for $K^c$ reduces to $K^c\rightarrow \frac{4}{F^2}g_{id}y^i\frac{\partial \mathcal{L}}{\partial g_{cd}}$ and $\bar\partial_cK^c \rightarrow (\frac{8}{F^4}g_{id}y^ig_{jc}y^j+\frac{4}{F^2}g_{cd})\frac{\partial \mathcal{L}}{\partial g_{cd}}$. Collecting all terms in the variation of the matter action in the metric geometry limit finally yields
\begin{equation}
\delta S_m[L,\Phi] \rightarrow \int_\Sigma d^4xd^3u\ \sqrt{g^Fh^F}_{|\Sigma} \Big[4\mathcal{L}-4g_{cd}\frac{\partial \mathcal{L}}{\partial g_{cd}}-24y^cy^d\frac{\partial \mathcal{L}}{\partial g_{cd}}\Big]_{|\Sigma}\frac{\delta L}{nL}\,,
\end{equation}
from which we can read off the expression for the source term $T_{|\Sigma}$,
\begin{equation}
T_{|\Sigma} \rightarrow \Big(4\mathcal{L}-4g_{cd}\frac{\partial \mathcal{L}}{\partial g_{cd}}-24y^cy^d\frac{\partial \mathcal{L}}{\partial g_{cd}}\Big)_{|\Sigma}\,.
\end{equation}
The lift of this expression to $TM$ requires making all terms zero homogeneous by multiplication with the appropriate powers of $F(x,y)$, which here means multiplication of the third term by $F(x,y)^{-2}$. The result confirms equation~(\ref{eq:tsig}) that was used to prove the consistency of Finsler gravity with Einstein gravity in the metric geometry limit.

\subsection{Complete lifts of cosmological symmetry generators}\label{app:compl}
We deduced the most general fundamental geometry function $L$ for Finsler spacetimes with cosmological symmetries  in section~\ref{sec:fsym}. The derivation requires the complete lifts of the symmetry-generating vector fields~(\ref{eq:csym}) which we display here explicitly:
\begin{eqnarray}
X^C_1&=&\chi\Big(\sin\theta \cos\phi
\partial_r+\frac{\chi}{r}\cos\theta\cos\phi\partial_\theta-\frac{\chi}{r}\frac{\sin\phi}{\sin\theta}\partial_\phi\Big)\nonumber\\
&&{}+\left(y^r\chi'\sin\theta\cos\phi+y^\theta\xi\cos\theta\cos\phi-y^\phi\xi\sin\theta\sin\phi\right)\bar\partial_r\nonumber\\
&&{}+\Big(y^r\big(\frac{\chi}{r}\big)'\cos\theta\cos\phi-y^\theta\frac{\chi}{r}\sin\theta\cos\phi-y^\phi\frac{\chi}{r}\cos\theta\sin\phi\Big)\bar\partial_\theta\\
&&{}+\Big(-y^r\big(\frac{\chi}{r}\big)'\frac{\sin\phi}{\sin\theta}+y^\theta\frac{\chi}{r}\frac{\sin\phi}{\sin^2\theta}\cos\theta-y^\phi\frac{\chi}{r}
\frac{\cos\phi}{\sin\theta}\Big)\bar\partial_\phi\nonumber\,,
\end{eqnarray}
\begin{eqnarray}
X^C_2&=&\chi\sin\theta \sin\phi
\partial_r+\frac{\chi}{r}\cos\theta\sin\phi\partial_\theta+\frac{\chi}{r}\frac{\cos\phi}{\sin\theta}\partial_\phi\nonumber\\
&&{}+\left(y^r\chi'\sin\theta\sin\phi+y^\theta\xi\cos\theta\sin\phi+y^\phi\xi\sin\theta\cos\phi\right)\bar\partial_r\nonumber\\
&&{}+\Big(y^r\big(\frac{\chi}{r}\big)'\cos\theta\cos\phi-y^\theta\frac{\chi}{r}\sin\theta\sin\phi+y^\phi\frac{\chi}{r}\cos\theta\cos\phi\Big)\bar\partial_\theta\\
&&{}+\Big(y^r\big(\frac{\chi}{r}\big)'\frac{\cos\phi}{\sin\theta}-y^\theta\frac{\chi}{r}\frac{\cos\phi}{\sin^2\theta}\cos\theta-y^\phi\frac{\chi}{r}\frac{\sin\phi}{\sin\theta}\Big)\bar\partial_\phi\nonumber\,,
\end{eqnarray}
\begin{equation}
X^C_3 = \chi\cos\theta
\partial_r-\frac{\chi}{r}\sin\theta\partial_\theta+\Big(y^r\chi'\cos\theta-y^\theta\chi\sin\theta\Big)\bar\partial_r-\Big(y^r\big(\frac{\chi}{r}\big)'\sin\theta+y^\theta\frac{\chi}{r}\cos\theta\Big)\bar\partial_\theta\,.
\end{equation}
The complete lifts $X_4^C$, $X_5^C$ and $X_6^C$ are stated in equations~(\ref{eq:ssymc}). In the formulae above we use the abbreviation $\chi=\sqrt{1-kr^2}$ and primes denote differentiation with respect to the coordinate $r$.

\subsection{Linearization identities}\label{app:lin}
In order to study the Finsler gravitational field equation perturbatively we have considered a class of Finsler spacetimes that are mild deviations from metric geometry in section~\ref{subsec:ModLGeom}. Here we list for completeness how to rewrite the appearing geometric objects in terms of the perturbation $h$ instead of the variable $l$ used in the main text.  

First we rewrite various derivatives acting on $l$ in terms of derivatives acting on $h$: 
\begin{subequations}
\begin{eqnarray}
 \nabla_a\nabla_b l&=&\frac{G^{1-k}}{k}\nabla_a\nabla_b h\,,\\
\bar\partial_a l&=&\frac{G^{1-k}}{k}\bar\partial_a h+\frac{(1-k)G^{-k}}{k}2 g_{ai}y^ih\,,\\
\nabla_a\nabla \bar\partial_q l&=&\frac{G^{1-k}}{k}\nabla_a\nabla \bar\partial_q h+\frac{2(1-k)G^{-k}}{k} g_{qi}y^i\nabla_a\nabla h\,,\\
\bar\partial_a\bar\partial_b l&=&\frac{G^{1-k}}{k}\bar\partial_a\bar\partial_b h+\frac{2(1-k)G^{-k}}{k}\Big(g_{bi}y^i\bar\partial_a
h+g_{ai}y^i\bar\partial_bh+(g_{ab}-\frac{2k}{G}g_{ai}y^ig_{bj}y^j)h\Big),\\
\nabla\nabla\bar\partial_a\bar\partial_b l&=&\frac{G^{1-k}}{k}\nabla\nabla\bar\partial_a\bar\partial_b
h\nonumber\\
&&{}+\frac{2(1-k)G^{-k}}{k}\nabla\nabla\Big(g_{bi}y^i\bar\partial_a h+g_{ai}y^i\bar\partial_bh+(g_{ab}-\frac{2k}{G}g_{ai}y^ig_{bj}y^j) h\Big),\\
g^{ab}l_{ab}&=&\frac{1}{2}g^{ab}\bar\partial_a\bar\partial_b l=\frac{G^{1-k}}{2k}g^{ab}\bar\partial_a\bar\partial_b h+\frac{2(1-k)(k+2)}{k}G^{-k}h\,,\\
\nabla_qg^{ab}l_{ab}&=&\frac{1}{2}g^{ab}\nabla_q\bar\partial_a\bar\partial_b l=\frac{G^{1-k}}{2k}\nabla_q g^{ab}\bar\partial_a\bar\partial_b
h+\frac{2(1-k)(k+2)}{k}G^{-k}\nabla_qh\,.
\end{eqnarray}
\end{subequations}

These identities can now be employed to determine the curvature scalar $\mathcal{R}=R^a{}_{ab}y^b$ of the Cartan non-linear connection which we use as the basic ingredient in our construction of Finsler gravity:
\begin{eqnarray}
 R^a{}_{ab}y^b&=&-y^ay^bR_{ab}[g(x)]-\frac{G^{1-k}}{2k}g^{ag}\big(\nabla_a\nabla \bar\partial_q h-\nabla_a\nabla_q h-\frac{1}{2}\nabla\nabla
\bar\partial_a\bar\partial_q h\big)\\
&&{}-\frac{2(1-k)G^{-k}}{2k}g^{aq}\big(g_{qi}y^i\nabla_a\nabla h-\frac{1}{2}\nabla\nabla\big(g_{qi}y^i\bar\partial_a
h+g_{ai}y^i\bar\partial_qh+(g_{aq}-\frac{2k}{G}g_{ai}y^ig_{qj}y^j) h\big)\big)\nonumber\\
&=&-y^ay^bR_{ab}[g(x)]-\frac{G^{1-k}}{2k}g^{ag}\big(\nabla_a\nabla \bar\partial_q h-\nabla_a\nabla_q h-\frac{1}{2}\nabla\nabla \bar\partial_a\bar\partial_q
h\big)+\frac{(1-k)(1+k)G^{-k}}{k}\nabla\nabla h\,.\nonumber
\end{eqnarray}

Finally we display how to rewrite the d-tensor $S$ and various derivatives acting on it; using the notation $\textrm{tr }h=g^{ab}\bar\partial_a\bar\partial_b h$ we find:
\begin{subequations}
\begin{eqnarray}
 S_p&=&-\frac{G^{1-k}}{4k}y^q\bar\partial_p\nabla_q (\textrm{tr }h)-\frac{(1-k)(k+2)}{k}G^{-k}y^q\bar\partial_p\nabla_qh\nonumber\\ 
&&{}-\frac{(1-k)G^{-k}}{2k}g_{pi}y^i\nabla (\textrm{tr }h)+2(1-k)(k+2)G^{-(1+k)}g_{pi}y^i\nabla h\,,\\
 \nabla_aS_b&=&-\frac{G^{1-k}}{4k}y^q\nabla_a\bar\partial_b\nabla_q \textrm{tr }h-\frac{(1-k)(k+2)}{k}G^{-k}y^q\nabla_a\bar\partial_b\nabla_qh\nonumber\\
&&{}-\frac{(1-k)G^{-k}}{2k}g_{bi}y^i\nabla_a\nabla (\textrm{tr }h)+2(1-k)(k+2)G^{-(1+k)}g_{bi}y^i\nabla_a\nabla h\,,\\
g^{ab}\nabla_aS_b&=&-\frac{G^{1-k}}{4k}y^qg^{ab}\nabla_a\bar\partial_b\nabla_q
(\textrm{tr }h)-\frac{(1-k)(k+2)}{k}G^{-k}y^qg^{ab}\nabla_a\bar\partial_b\nabla_qh\nonumber\\
&&{}-\frac{(1-k)G^{-k}}{2k}\nabla\nabla (\textrm{tr }h)+2(1-k)(k+2)G^{-(1+k)}\nabla\nabla h\,,
\end{eqnarray}
\begin{eqnarray}
g^{ab}y^p\bar\partial_a\nabla_pS_b&=&-\frac{(1-k)(2+k)}{k}G^{-k}\Big(y^qy^pg^{ab}\bar\partial_a\nabla_p\bar\partial_b\nabla_q
h+g^{ab}\nabla\bar\partial_a\nabla_b h\Big)\nonumber\\
&&{}-\frac{G^{1-k}}{4k}\Big(y^qy^pg^{ab}\bar\partial_a\nabla_p\bar\partial_b\nabla_q (\textrm{tr }h)+g^{ab}\nabla\bar\partial_a\nabla_b (\textrm{tr }h)\Big)\\
&&{}+2(1-k)(2+k)(2k+3)G^{-(k+1)}\nabla\nabla h-\frac{(1-k)(2k+1)}{2k}G^{-k}\nabla\nabla (\textrm{tr }h)\,.\nonumber
\end{eqnarray}
\end{subequations}


\end{document}